\documentclass[aps,prb,a4paper,twocolumn,10pt,floatfix,showpacs,showkeys,amsmath,amssymb,nobibnotes,unsortedaddress,superscriptaddress]{revtex4-2}

\setcitestyle{super}
\usepackage{graphicx}
\usepackage[export]{adjustbox}
\usepackage[dvipsnames]{xcolor}
\usepackage[utf8]{inputenc}
\usepackage[T1]{fontenc}
\usepackage{amsmath}
\usepackage{xr}
\usepackage{upgreek}
\usepackage[percent]{overpic}
\usepackage{xcolor}
\usepackage{physics}
\usepackage{verbatim}
\usepackage[explicit]{titlesec}
\usepackage{lipsum}
\usepackage{xspace}

\graphicspath{{figures/}}

\renewcommand{\thesection}{\arabic{section}}

\titleformat{\section}{\large\bfseries\filcenter}{\thesection.}{1em}{#1}

\newcommand{\jgen}{j_\mathrm{gen}}

\newcommand{\jph}{j_\mathrm{photo}}
\newcommand{\jsc}{j_\mathrm{sc}}

\newcommand{\Voc}{V_\mathrm{oc}}
\newcommand{\Vimp}{V_\mathrm{imp}}
\newcommand{\Vext}{V_\mathrm{ext}}

\newcommand{\Vtr}{\Delta V_\mathrm{tr}}
\newcommand{\FF}{F\!F}
\newcommand{\pFF}{pF\!F}

\newcommand{\rext}{R_\mathrm{ext}}
\newcommand{\nid}{n_\mathrm{id}}
\newcommand{\nsig}{n_\mathrm{\sigma}}

\newcommand{\napp}{n_\mathrm{app}}

\newcommand{\eg}{E_{g}}
\newcommand{\etacol}{\eta_\mathrm{col}}
\newcommand{\etarec}{\eta_\mathrm{rec}}
\newcommand{\bl}{\left(}
\newcommand{\bL}{\left[}
\newcommand{\br}{\right)}
\newcommand{\bR}{\right]}

\newcommand{\kT}{k_B T}
\newcommand{\voc}{v_\mathrm{oc}}

\newcommand{\vmpp}{v_\mathrm{mpp}}

\begin{document}

\title{Transport resistance strikes back: unveiling its impact on fill factor losses in organic solar cells}

\author{Maria Saladina}
\email[email: ]{maria.saladina@physik.tu-chemnitz.de}
\affiliation{Institut für Physik, Technische Universität Chemnitz, 09126 Chemnitz, Germany}

\author{Carsten Deibel}
\email[email: ]{deibel@physik.tu-chemnitz.de}
\affiliation{Institut für Physik, Technische Universität Chemnitz, 09126 Chemnitz, Germany}

\begin{abstract}

The fill factor ($\FF$) is a critical parameter for solar cell efficiency, but its analytical description is challenging due to the interplay between recombination and charge extraction processes. An often overlooked yet significant factor contributing to $\FF$ losses, beyond recombination, is the influence of charge transport. In most state-of-the-art organic solar cells, the primary limitations of the $\FF$ arise not just from non-radiative recombination but also from low conductivity. A closer look reveals that even in the highest efficiency cells, performance losses due to transport resistance are significant, highlighting the need for refined models to predict the $\FF$ accurately. 

Here, we extend the analytical model for transport resistance to a more general case. Drawing from a large set of experimental current--voltage and light intensity-dependent open-circuit voltage data, we systematically incorporate crucial details previously omitted in the model. Consequently, we introduce a straightforward set of equations to predict the $\FF$ of a solar cell, enabling the differentiation of losses attributed to recombination and transport resistance. Our study provides valuable insights into strategies for mitigating $\FF$ losses based on the experimentally validated analytical model, guiding the development of more efficient solar cell designs and optimisation strategies.

\end{abstract}


\maketitle

\section{Introduction}

Lately, we have witnessed remarkable progress in enhancing the efficiency of organic solar cells (OSCs),\cite{wang2023binary,deng2023type} paving the way for financially viable upscaling. Solar cells with thicker active layers may be more compatible with printing or coating processes.\cite{wachsmuth2023fullyprinted,zhang2024sequentially} This requires optimising charge collection efficiency and reducing transport resistance losses, both of which become more critical with increased thickness.\cite{lubke2023understanding,schiefer2014determination} Even lab-scale record efficiency devices with optimised active layer thickness suffer from charge carrier collection losses. For example, the $\FF$ of a single-junction OSC with an average power conversion efficiency of 19.3\% stands at 79.6\%.\cite{liu2024recordPCE} Based on the provided data, we estimate that the $\FF$ would reach 87.4\% if transport resistance losses were eliminated. The difference of 7.8 percentage points is direct evidence that transport resistance is a relevant loss mechanism, even in these record-efficiency solar cells. 

Among the main loss mechanisms in OSCs, transport resistance has been overlooked compared to the more extensively studied geminate and nongeminate recombination.\cite{gohler2018nongeminate,shoaee2019decoding,bronstein2020role,saladina2021charge} Several studies predicted that, besides recombination, the reduction in $\FF$ of OSCs was attributed to slow charge carrier transport,\cite{cheyns2008analytical,stelzl2012modeling} and depended on the ratio of charge carrier recombination and extraction rates.\cite{bartesaghi2015competition} Würfel et al.\ demonstrated using drift--diffusion simulations that slow charge carrier transport led to the accumulation of charge carriers within the device.\cite{wurfel2015impact} This accumulation caused a substantial difference between the applied voltage considered in the diode equation and the actual quasi-Fermi level splitting (QFLS), leading to a gradient of the quasi-Fermi levels in the active layer. Neher et al.\ showed how the slope of the current--voltage ($j(V)$) curve around the open-circuit voltage ($\Voc$) influences the $\FF$ and alters the apparent ideality factor in the diode equation.\cite{neher2016new} The slope was parameterised by the figure of merit $\alpha$, a measure of transport-induced series resistance near $\Voc$. A figure of merit reciprocal to $\alpha$ was related to the $\FF$ by Kaienburg et al.,\cite{kaienburg2016extracting} and by Xiao et al.\cite{xiao2020relationship} with focus on the impact of tail states.

The validity of these predictions is now supported by experimental data, as $\FF$ losses have been attributed to transport resistance in various OSCs.\cite{tokmoldin2021explaining} In our study on thermal degradation in PM6:Y6 solar cells, transport resistance was related to device stability.\cite{wopke2022traps} We identified transport resistance, likely originating from increased defect formation, as the primary factor driving the drop in photovoltaic performance. A positive effect on both $\FF$ and solar cell stability was observed by Yang et al.\ in doped OSCs, where they noted improved charge carrier collection efficiency.\cite{yang2019black} The influence of transport resistance on $\FF$ extends beyond OSCs,\cite{rau2020luminescence,grabowski2022fill} indicating collection efficiency losses even in semiconductors with comparably higher mobilities. 

Despite clear evidence of its detrimental impact on solar cell performance, experimental measurements of transport resistance remain scarce. While models of transport resistance do provide qualitative predictions for the $\FF$, they neglect trap states entirely and can predict only rough trends observed in measurements.

In this paper, we refine the analytical model for transport resistance based on experimental data. We extend the diode equation to accommodate both limited extraction and recombination, considering trapping and allowing for non-unity ideality factors. We establish a precise method for evaluating the effective conductivity at open circuit through straightforward measurements of $j(V)$ and the light intensity-dependent $\Voc$ of a solar cell. Our model allows to understand the contributions of the two charge carrier types to recombination and transport, and to predict the $\FF$ limitations at both open circuit and the maximum power point (MPP). Based on the experimentally verified analytical model, our study provides insights into approaches aimed at minimising $\FF$ losses.

\section{Results and Discussion}

\subsection{The current--voltage characteristics are transport resistance limited, the suns-$\Voc$ curve is not.}

\begin{figure}[b]\centering
    \includegraphics[width=0.38\textwidth]{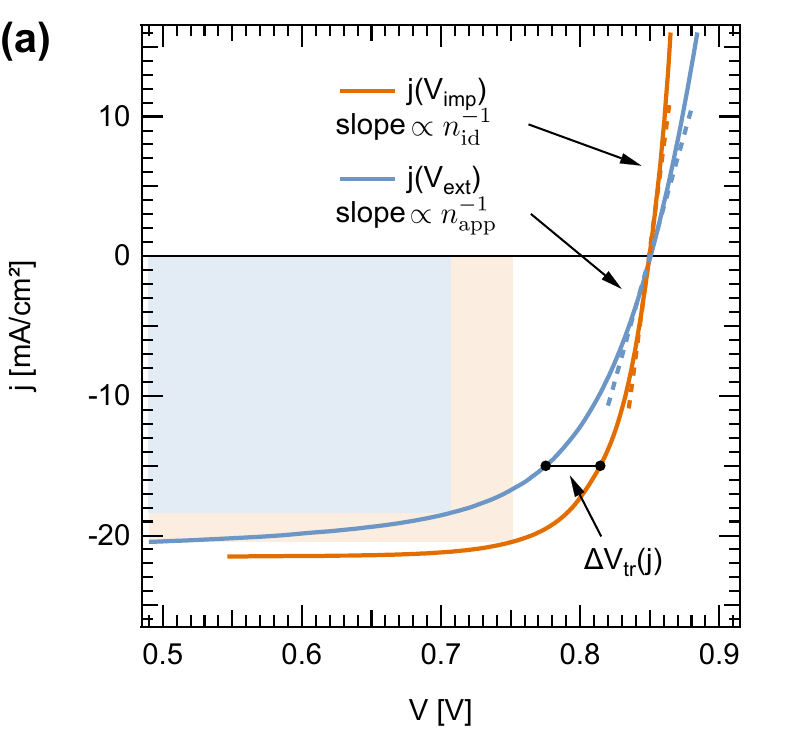}
    \\[\baselineskip]
    \includegraphics[width=0.38\textwidth]{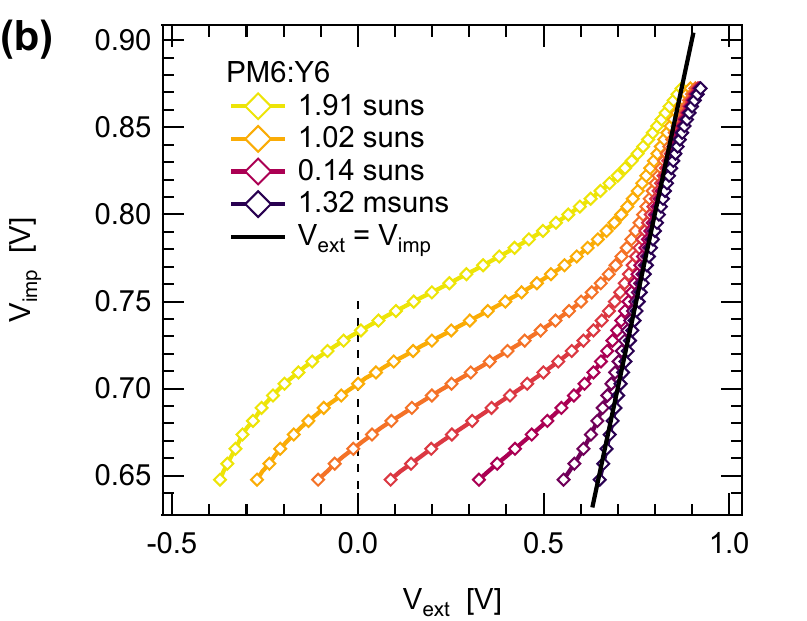}
    \caption{(a) A suns-$\Voc$ curve, $j(\Vimp)$, contains recombination losses, while an illuminated $j(V)$ curve, $j(\Vext)$, incorporates additional losses due to transport resistance. Their slopes at $\Voc$ are inversely proportional to $\nid$ and $\napp$, respectively. The shaded areas correspond to the output power at their respective MPPs. The difference between $\Vimp$ and $\Vext$ at a given current density is the voltage drop due to transport resistance, $\Vtr$. (b) The difference between $\Vimp$ and $\Vext$ represents a severe loss at higher light intensities. Notably, under short circuit (dashed line) the solar cell is effectively operating close to $\Voc$. }
\label{fig:01}\end{figure}

Transport resistance is an internal resistance within the active and transport layers of a solar cell resulting from relatively slow movement of charge carriers, effectively acting as an internal series resistance. As shown in Figure~\ref{fig:01}(a), we evaluate this resistance by comparing an illuminated current--voltage curve, $j(\Vext)$, which encompasses both recombination and transport resistance losses, to its resistance-free counterpart, $j(\Vimp)$. The latter is estimated by a suns-$\Voc$ curve, i.e.\ the open-circuit voltage of a solar cell measured over several orders of magnitude of light intensity shifted by the generation current density, using a previously established method.\cite{muller2013analysis,schiefer2014determination,mackel2018determination,wopke2022traps} In this context, $\Vimp$ represents the open-circuit voltage that is measured under a specific light intensity.\cite{faisst2022direct,list2023determination} For more details please refer to Section~S2 in the Supporting Information.

The suns-$\Voc$ curve accounts solely for recombination losses, excluding transport resistance. Considering that the net current is a result of generation and recombination currents in a solar cell, it can be characterised analytically using the diode equation\cite{wurfel2015impact,neher2016new}
\begin{equation}\begin{split}\label{eq:JVimp}
    j(\Vimp)
    &= j_0 \cdot \exp\bl \frac{e\Vimp}{\nid\kT} \br - \jgen \\
    &= \jgen \left[ \exp\bl \frac{e\bl\Vimp - \Voc\br}{\nid \kT} \br - 1 \right] , \\
\end{split}\end{equation}
where $\jgen$ stands for the total generation current density, the sum of the dark saturation current density $j_0$ and the photocurrent density $\jph$, $\nid$ stands for the recombination ideality factor, $k_B$ the Boltzmann constant, and $T$ the temperature (see Figure~\ref{fig:01}(a)). The above equation is equivalent to the well-known ideal diode equation. We used the relation $\jgen = j_0 \exp\bl e\Voc/(\nid\kT) \br$, solved for $j_0$, to change the familiar form of the equation, which becomes useful later when we focus on the slope of the $j(V)$ curve. We emphasise that the suns-$\Voc$ curve depends on the implied voltage $\Vimp$, i.e.\ the QFLS divided by the elementary charge $e$.

For infinite charge carrier mobility, an externally applied voltage $\Vext$ equals $\Vimp$, the voltage that charge carriers feel within the device: the quasi-Fermi levels in the bulk are flat and the transport resistance is zero. However, in reality, charge carrier mobility is finite. Slow charge carrier transport leads to a tilting of the quasi-Fermi levels and gives rise to a discrepancy between $\Vext$ and $\Vimp$,\cite{wurfel2015impact} as shown in Figure~\ref{fig:01}(b) for PM6:Y6. This discrepancy is what we refer to as the \emph{voltage loss due to transport resistance}, defined as\cite{schiefer2014determination,neher2016new}
\begin{equation}\begin{split}\label{eq:Vtr}
    \Vtr &= j \cdot \frac{L}{\sigma} = \frac{\nabla E_F}{e}\cdot L .
\end{split}\end{equation}
Here, $j$ is the current density, $L$ the active layer thickness, $\sigma$ the conductivity, and $\nabla E_F$ the gradient of the quasi-Fermi levels, which contains all contributions of non-equilibrium on charge transport, including both drift and diffusion.\cite{nelson2003physics-book,sze2021physics} This voltage drop, due to the transport resistance and originating from a low conductivity, is what leads to the $\FF$ loss. The low conductivity can come from the active layer or the transport layers. The latter are often doped and generally more conductive than the active layer of the solar cell. Therefore, in the following we refer to the transport resistance with respect to the effective conductivity of the active layer. We determined $\Vtr$ from the experimental data as $\Vtr(j) = \Vext(j) - \Vimp(j) - j\rext$, i.e.\ the difference between the illuminated $j(V)$ and the suns-$\Voc$ curve at the same current density, while also factoring in $\rext$, the series resistance of the circuit.\cite{schiefer2014determination} The latter was estimated from fitting $d\Vext/dj$ at high forward bias, where transport resistance is negligible (see Figure~S2).

Because of the discrepancy between $\Vext$ and $\Vimp$, the ideal diode equation is not suitable to describe the illuminated $j(V)$ curve. Considering the above relation between $\Vext$ and $\Vimp$ at the same $j$, Eq.~\eqref{eq:JVimp} can be modified to account for the influence of transport and external series resistance,
\begin{equation*}\begin{split}
    j(\Vext) &= \jgen \left[ \exp\bl \frac{e\bl\Vext - \Vtr - j\rext-\Voc\br}{\nid \kT} \br - 1 \right] . 
\end{split}\end{equation*}
We note that the effect of $\rext$ in the fourth quadrant of the $j(\Vext)$ curve is negligible (see Figure~S3), and while it is considered in determination of $\Vtr$ from the experimental data, further analytical treatment assumes that $\rext\approx0$ for clarity. 

For an easier comparison between the illuminated $j(V)$ curve and the series resistance free suns-$\Voc$ curve, we define an apparent ideality factor for the illuminated $j(V)$ curve as $\napp=\nid+\beta$, where
\begin{equation}\begin{split}\label{eq:beta}
    \beta(\Vext) &= \nid\cdot\frac{\Vtr}{\Vimp-\Voc} . 
\end{split}\end{equation}
This ideality factor, which accounts for the combined influence of recombination and transport resistance, allows for a simple expression of $j(\Vext)$ that can be more easily compared to Eq.~\eqref{eq:JVimp}:
\begin{equation}\begin{split}\label{eq:JVext}
    j(\Vext) 
    &= \jgen \left[ \exp\bl \frac{e\bl\Vext - \Voc\br}{\napp \kT} \br - 1 \right] .
\end{split}\end{equation}
We emphasise that Eqs.~\eqref{eq:JVimp} and \eqref{eq:JVext} are equivalent when the corresponding voltages are evaluated at the same current density. 

The above equation describes the $j(V)$ curve of a typical state-of-the-art organic solar cell in the fourth quadrant. Hopping transport is represented by the effective conductivity $\sigma$, which enters via $\Vtr$. The impact of energetic disorder on recombination is included via the ideality factor $\nid$. We note that extraction barriers or leakage currents due to low shunt resistance are not considered, as we do not observe that these losses play a role in our devices. Instead, Eq.~\eqref{eq:JVext} encompasses the dominant loss mechanisms in state-of-the-art OSCs: nongeminate recombination and transport resistance. 

Charge generation in Eq.~\eqref{eq:JVext} is assumed to be voltage-independent, since geminate recombination is not the dominant loss mechanism in state-of-the-art OSCs \cite{perdigon2020barrierless}. However, Eq.~\eqref{eq:JVext} can be easily extended to the case where $\jgen$ depends on the externally applied electric field, albeit at the expense of simplicity. For the reassurance of the reader and ourselves, we have dedicated Section~S4 in the Supporting Information to address the potential impact of voltage-dependent charge generation on the current--voltage characteristics and the fill factor. The results indicate that $\Vtr$ remains the primary fill factor loss, for devices in which the generation yield at zero field is at least half of the precursor states.

We will evaluate transport resistance losses by examining the apparent ideality factor of the illuminated $j(V)$ curve. The ideality factor changes the slope of the $j(V)$ curve, thus affecting both the MPP and the $\FF$. The resistance-free suns-$\Voc$ curve in Figure~\ref{fig:01}(a) is solely impacted by recombination, and its slope is inversely proportional to $\nid$, as follows from Eq.~\eqref{eq:JVimp}. The ideality factor of the illuminated $j(V)$ curve, $\napp$, is increased by the additional term $\beta$ containing the transport resistance loss $\Vtr$. This alteration results in a shallower slope and causes the operating point to deviate from the theoretically achievable MPP, leading to a lower $\FF$.

\subsection{The fill factor is almost always more limited by transport resistance than recombination.}

The $\FF$ is a key parameter in determining the overall efficiency of a solar cell. We can differentiate between the overall $\FF$ of the current--voltage characteristics under illumination and the higher pseudo-fill factor, denoted as $\pFF$, that only represents recombination losses. The $\pFF$, therefore, characterises the fill factor of a solar cell when transport and external series resistance are absent and represents its higher limit.\cite{green1981solar,wopke2022traps,grabowski2022fill} 

\begin{figure}[tb]\centering
    \includegraphics[width=0.38\textwidth]{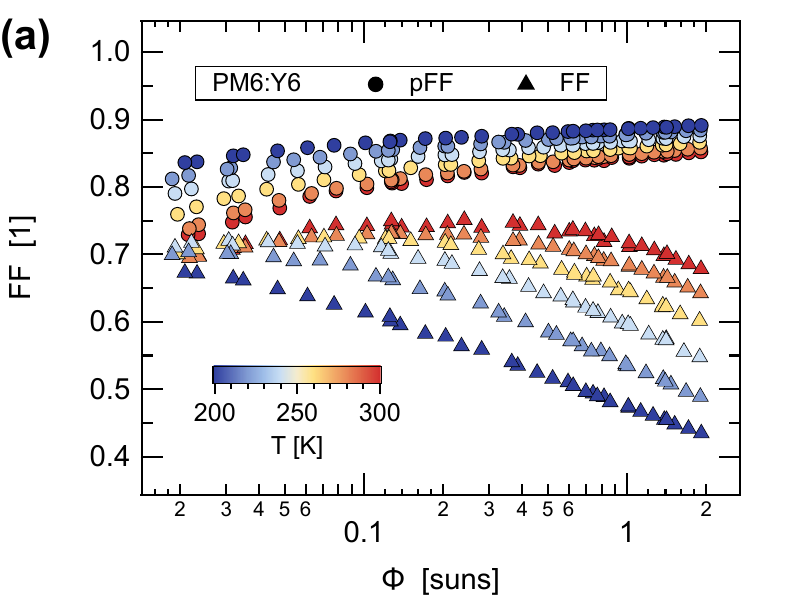}
    \\[\baselineskip]
    \includegraphics[width=0.38\textwidth]{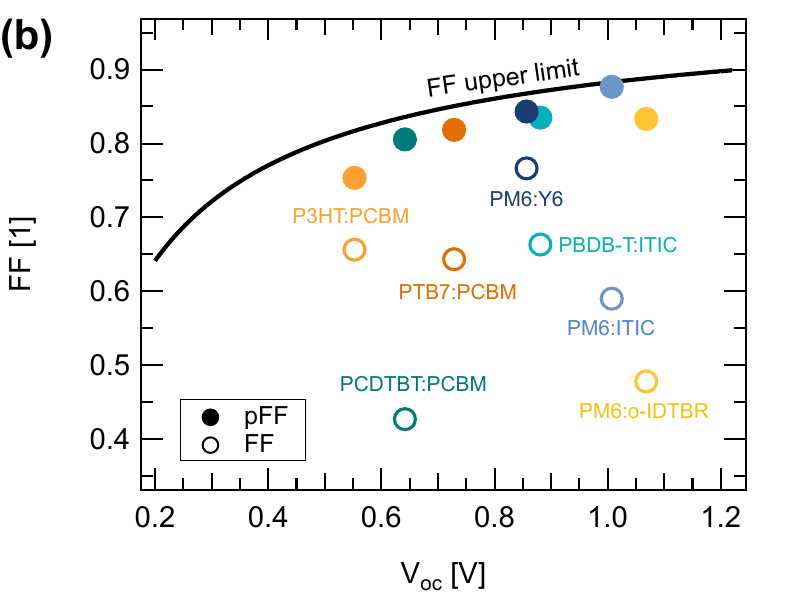}
    \caption{(a) $\pFF$ and $\FF$ of a PM6:Y6 solar cell. While $\pFF$ increases with light intensity $\Phi$ owing to reduced disorder, $\FF$ decreases due to higher $\Vtr$. (b) Comparison of $\pFF$ and $\FF$ in OSCs based on different donor--acceptor blends. The $\FF$ upper limit (solid line) was determined using Eqs.\eqref{eq:FFvnorm} and ~\eqref{eq:FFgreen} with $\napp=1$ and $m=0.72$. The difference between the ideal line and $\pFF$ being smaller than the one from $\FF$ to $\pFF$ represents the dominant limitation by transport resistance.}
\label{fig:02}\end{figure}

Figure~\ref{fig:02}(a) presents a comparison of the fill factors for a PM6:Y6 solar cell obtained from $j(\Vimp)$ (Eq.~\eqref{eq:JVimp}), determined by the suns-$\Voc$ method, and $j(\Vext)$ (Eq.~\eqref{eq:JVext}), which were measured over a broad range of temperatures and light intensities.
The $\pFF$ tends to improve with higher light intensity, which we attribute to reduced energetic disorder as the QFLS increases.\cite{saladina2023power} In contrast, the $\FF$ of an illuminated $j(V)$ curve tends to decrease with increasing light intensity, as the transport resistance loss $\Vtr$ becomes more significant. For the PM6:Y6 solar cells and other systems summarised in Figure~\ref{fig:02}(b), the losses due to transport resistance outweigh those caused by recombination. Almost always, voltage loss due to transport resistance is the primary contributor to $\FF$ losses in OSCs under operating conditions.

We will examine in more detail how transport resistance impacts the $\FF$ due to effective conductivity. First, we will consider the slope of the illuminated $j(V)$ curve around $\Voc$: it is directly related to effective conductivity. We will then extend this analysis to the MPP, where the $\FF$ is defined, in order to get a unified view of the impact of transport resistance on the $\FF$. We will conclude with our perspective on how the transport resistance can be minimised and, thus, the $\FF$ optimised.

\subsection{The open circuit: The effective conductivity and the figure of merit $\alpha$.}

The slope of the illuminated $j(V)$ curve is reduced due to transport resistance. We accounted for this dependence by an apparent ideality factor $\napp=\nid+\beta$. The impact of the transport resistance is expressed by the parameter $\beta$, which we defined in Eq.~\eqref{eq:beta}. In a later section, we will use this parameter $\beta$ to present a generalised description of the $\FF$ and the impact of transport resistance.

Previously, Neher et al.\ showed that near open circuit $\napp = 1 + \alpha$, where $\nid = 1$ and $\alpha$ was the figure of merit for OSCs with transport-limited photocurrents.\cite{neher2016new} In our notation, $\alpha = \beta(\Voc)$. We generalise the valuable insight by the Neher group, as OSCs are generally energetically disordered with charge transport and recombination dominated by energetic traps: in most cases, $\nid$ differs from unity, and we find that the apparent ideality factor becomes $\nid+\alpha$ (for derivation please refer to Section~S5 in the Supporting Information). This correction becomes important when the values of $\alpha$ are comparable to $\nid$. Essentially, $\alpha$ is a measure of the competition between recombination and conductivity at open circuit conditions: 
\begin{equation}\begin{split}\label{eq:alpha_short}
    \alpha &= \frac{eL}{\kT} \cdot \frac{\jgen}{\sigma_{\Voc}} , \\
\end{split}\end{equation}
where the recombination current density equals $\jgen$ at $\Voc$, and $\sigma_{\Voc}$ is the effective conductivity at open circuit. To gain a precise understanding of the factors influencing $\alpha$, it is essential to comprehend the individual contributions of recombination and conductivity. Subsequently, our attention will be directed towards the latter, demonstrating how it can be assessed at $\Voc$ using only two datasets: the illuminated $j(V)$ curve and the transport resistance-free suns-$\Voc$ curve.

\subsubsection{The effective conductivity and the transport ideality factor.}
The effective conductivity can generally be determined from Eq.~\eqref{eq:Vtr}, although it leads to a discontinuity at $\Voc$, where $\Vtr = 0$.\cite{schiefer2014determination,mackel2018determination} We overcome this obstacle in a simple way using a derivative. As both $j$ and $\sigma$ change as we move along the $j(V)$ curve, the derivative $d\Vtr/d j$ given by Eq.~(S1) has two terms. At $\Voc$, however, only one term remains non-zero, leading to
\begin{equation}\begin{split}\label{eq:sigma_exp}
    \sigma_{\Voc} &= L\cdot \bl\left.\frac{d\Vtr}{d j}\right|_{j=0}\br^{-1} . \\
\end{split}\end{equation}
We apply the condition of equal electron and hole current densities in the bulk to Eq.~\eqref{eq:Vtr}, resulting in $\sigma_n \cdot \Delta V_\mathrm{tr,n} = \sigma_p  \cdot \Delta V_\mathrm{tr,p}$. If the conductivity of one charge carrier type is lower than the other, then $\Vtr$ is inevitably higher. Consequently, the $\FF$ is limited by the slower-moving charge carrier.

\begin{figure*}[htb]\centering
    \includegraphics[width=0.32\textwidth]{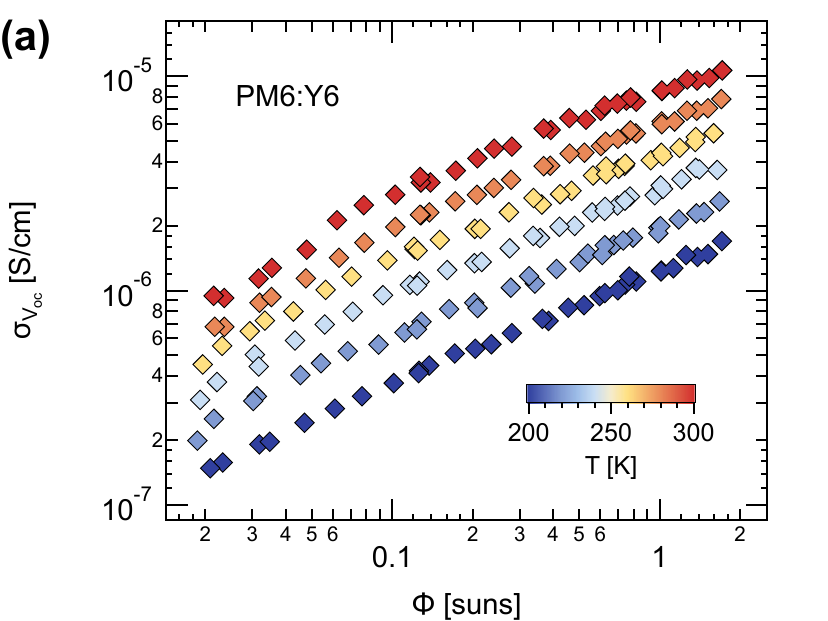}\quad
    \includegraphics[width=0.32\textwidth]{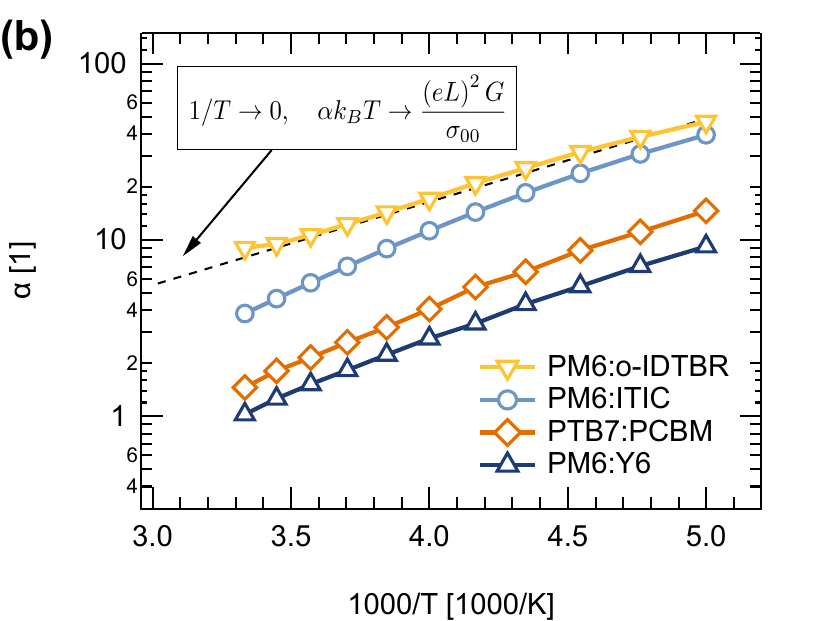}\quad
    \includegraphics[width=0.32\textwidth]{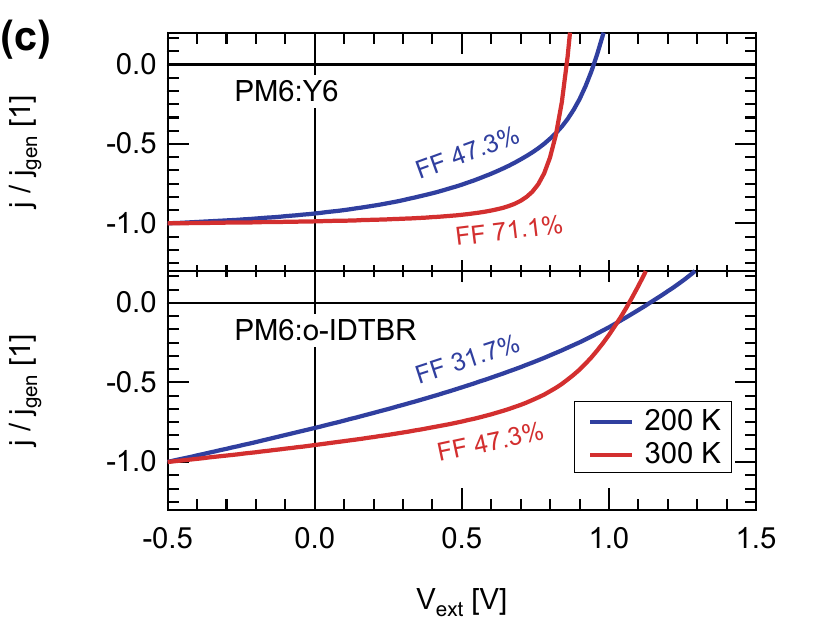}
    \caption{(a) $\sigma_{\Voc}$ determined from the slope of $\Vtr(j)$ according to Eq.~\eqref{eq:sigma_exp}. (b) Temperature dependence of $\alpha$ for different material systems. The magnitude of $\alpha$ at constant light intensity depends on $\sigma_{00}$. (c) Comparison of the $j(V)$ curves at 200~K and 300~K close to 1~sun. Higher $\alpha$ for PM6:o-IDTBR at 200~K severely impacts the $\FF$, leading to a transport-controlled $j(V)$ curve.}
\label{fig:03}\end{figure*}

The effective conductivity is determined from the slope of $\Vtr$ according to Eq.~\eqref{eq:sigma_exp}. The result is depicted in Figure~\ref{fig:03}(a) and shows that, as expected, the effective conductivity increases with higher light intensity $\Phi$ and temperature, following the rise in charge carrier density. Based on the multiple trapping and release model (see Section~S6 in the Supporting Information), we describe $\sigma$ as follows:
\begin{equation}\begin{split}\label{eq:sigma-T}
    \sigma_{\Voc} &= \sigma_{00} \cdot \exp\bl -\frac{\eg-e\Voc}{\nsig\kT}\br , \\
\end{split}\end{equation}
with $\sigma_{00}$ being a temperature-independent prefactor, and $\eg$ the effective energy gap. Analogous to the recombination ideality factor, the voltage dependence of conductivity is characterised by a transport ideality factor $\nsig$. While both ideality factors depend on the density of states (DOS) and how the ratio of mobile to all charge carriers depends on the QFLS, the contributions of electrons and holes are weighed differently. The recombination rate typically contains an arithmetic mean of electron and hole mobilities, according to the reduced Langevin model\cite{gohler2018nongeminate}, or a geometric mean for moderate donor--acceptor phase separation.\cite{heiber2015encounter} In contrast, $\sigma$ is dominated by the slower-moving charge carrier and is thus determined by the harmonic mean of electron and hole conductivities. 

In order to take a broader view of the impact of charge carrier transport, we consider four different organic solar cells: PM6:Y6, PM6:ITIC, PM6:o-IDTBR, and PTB7:PCBM (fabrication details are provided in Section~S1, Supporting Information). Their $\alpha$ values -- calculated from experimental data using Eq.~\eqref{eq:alpha_short} -- are shown in Figure~\ref{fig:03}(b) as a function of temperature. The data is evaluated under 1~sun illumination intensity. We assume that the charge carrier generation rate $G$ remains temperature-independent (or weakly temperature-dependent), so $\jgen = eLG$ can be considered roughly constant for each material system. The exponential temperature dependence of $\alpha$ in Eq.~\eqref{eq:alpha_short} is therefore primarily dictated by $\sigma$. As temperature rises, charge carrier hopping between localised sites becomes easier. In the multiple trapping and release model, this corresponds to a share of mobile charge carriers becoming larger, as they are more easily released from shallow traps to the transport energy level. Consequently, higher conductivity at a constant generation rate lowers the $\alpha$ value and enhances the $\FF$. 

If charge carriers had an infinite amount of thermal energy and could move freely within the device (i.e., the exponential term in Eq.~\eqref{eq:sigma-T} became 1), then $\alpha$ would be determined by the ratio between $G$ and $\sigma_{00}$. Among the four solar cells, this ratio is highest in PM6:o-IDTBR, as inferred from extrapolating the data to $1/T=0$. This explains why $\alpha$ is, on average, ten times higher than in a PM6:Y6 solar cell, which directly impacts the $\FF$ in Figure~\ref{fig:03}(c). PM6:Y6 at 200~K and PM6:o-IDTBR at 300~K both exhibit $\alpha\approx 9$, coinciding with identical fill factors at these temperatures. At 200~K, the $\alpha$ value for PM6:o-IDTBR increases fivefold, while the $\FF$ drops to a mere 31.7\%. However, even low values of $\alpha$ have a significant impact on the $\FF$. At 300K, $\alpha\approx 1$ for PM6:Y6, yet the fill factor in Figure~\ref{fig:03}(c) remains well below the $\pFF$ limit of 84\%. 

The situation differs when we consider $\sigma$ at the same \emph{temperature} and vary the \emph{generation rate}. Intuitively, we anticipate an improvement in $\FF$ with increased light intensity, as traps are filled and transport improves (at least this is true for an exponential distribution of trap states). However, contrary to this expectation, the data in Figure~\ref{fig:02}(a) shows that the $\FF$ of the PM6:Y6 solar cell decreases with higher illumination. The figure of merit $\alpha$ provides an explanation for this observation. As $\sigma$ in Eq.~\eqref{eq:alpha_short} increases, so does $\jgen$. Both of these competing processes depend on light intensity, yet $\jgen$ depends on it more strongly than $\sigma$, leading to an increase in $\alpha$ and an overall lower $\FF$. 

This work addresses the dominant $\FF$ losses in state-of-the-art OSCs, which are typically not limited by the electric field-dependence of charge generation. 
Among the four solar cells studied here, only PM6:o-IDTBR exhibits voltage-dependent charge generation.\cite{tokmoldin2023elucidating} 
For PM6:Y6, PM6:ITIC, and PTB7:PCBM, time-delayed collection field experiments demonstrated that this process is insensitive to the applied field.\cite{kniepert2015effect,perdigon2020barrierless,tokmoldin2023elucidating} 
However, the figure of merit $\alpha$ in PM6:o-IDTBR could, in principle, be influenced by charge generation. Figure~S4(d) shows that this effect is virtually absent even in the relatively bad solar cells, where the electron--hole pair dissociation efficiency is only 50\%.

\subsubsection{The impact of energetic disorder on the fill factor losses: Extending the figure of merit $\alpha$.}

\begin{figure}[htb]\centering
    \includegraphics[width=0.38\textwidth]{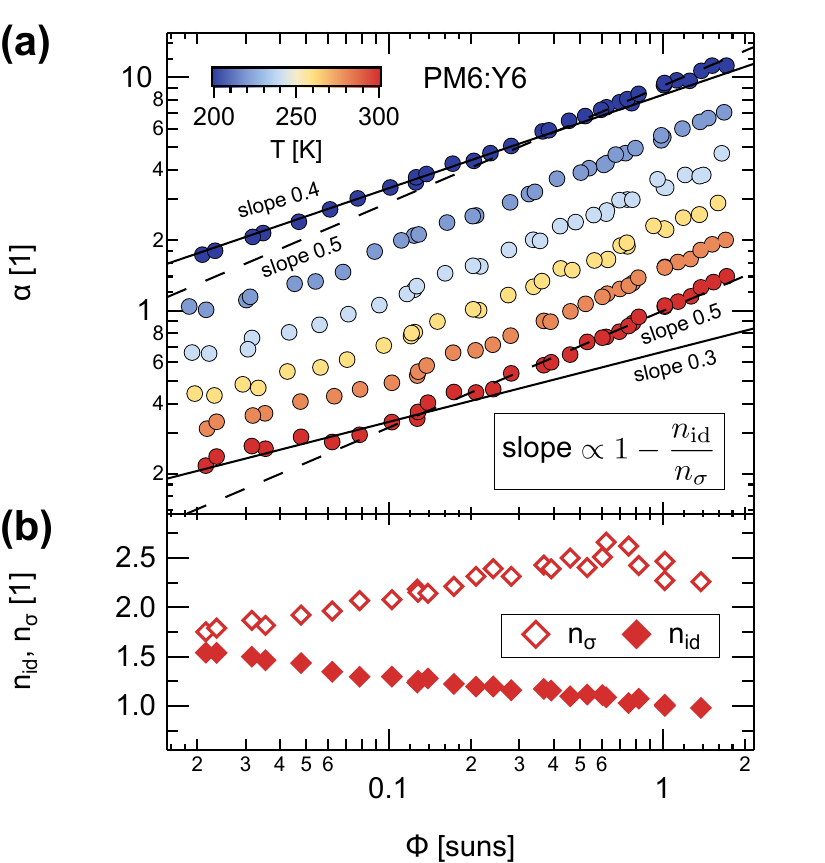}
    \\[\baselineskip]
    \includegraphics[width=0.38\textwidth]{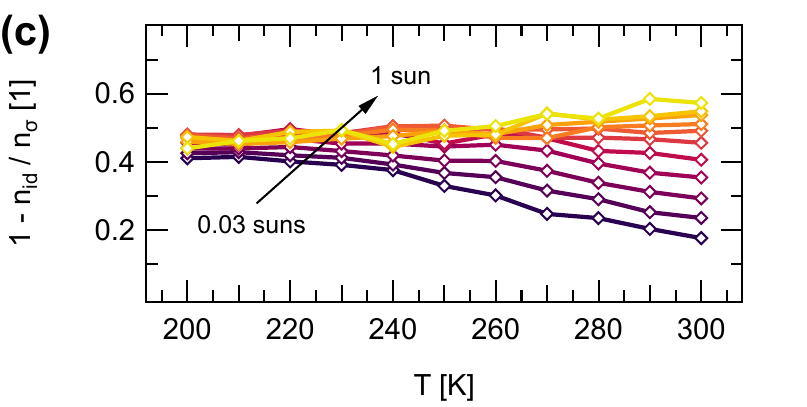}
    \caption{(a) Light intensity dependence of $\alpha$ for PM6:Y6. The slope at low $\Phi$ deviates from the commonly assumed $1/2$. (b) The ideality factors for recombination ($\nid$) and transport ($\nsig$) at 300~K. Deviations from the values of $\nid=1$ and $\nsig=2$, commonly assumed in the models, require incorporating these factors into the analytical expression for $\alpha$. (c) The term $1-\nid/\nsig$ is $\approx 0.5$ at higher $\Phi$ but decreases at lower $\Phi$, explaining the slope of $\alpha(\Phi)$.}
\label{fig:04}\end{figure}

To better understand how recombination and transport interact to influence the $\FF$, we examine the figure of merit $\alpha$ in more detail. Figure~\ref{fig:04}(a) presents the $\alpha$ values for a PM6:Y6 solar cell over a wide range of temperatures and light intensities. To understand the exact parameters affecting $\alpha$, we are interested in the slopes. The data demonstrates that $\alpha \propto \sqrt{\Phi}$, deviating at lower illumination intensities. From the definition of $\alpha$, Eq.~\eqref{eq:alpha_short}, it is evident that $\alpha$ is related to $\Phi$ through the light intensity dependence of both $\jgen$ and $\sigma_{\Voc}$. In previous models, $\alpha$ is typically assumed to be proportional to $\sqrt{\Phi}$, because $\jgen$ and $\sigma_{\Voc}$, in the simplest case (ignoring trap states or considering a Gaussian distribution of tail states in the low charge carrier concentration regime), scale with ideality factors of $\nid=1$ and $\nsig=2$, respectively.\cite{bartesaghi2015competition,neher2016new} However, this assumption is not accurate in the general case.\cite{foertig2012shockley,hawks2013relating,hofacker2017dispersive} 

Both $\jgen$ and $\sigma_{\Voc}$ have additional voltage dependencies. Generally, when the net current is zero,
\begin{equation*}\begin{split}\label{eq:jgen}
    \jgen = j_{00} \cdot \exp\bl -\frac{\eg-e\Voc}{\nid\kT}\br \propto \Phi , \\
\end{split}\end{equation*}
with $j_{00}$ denoting a temperature-independent prefactor.\cite{foertig2012shockley} The effective conductivity was already defined in Eq.~\eqref{eq:sigma-T}. The recombination ideality factor of PM6:Y6 in Figure~\ref{fig:04}(b) equals unity only within a narrow range of light intensities close to 1~sun. Similarly, $\sigma$ has a transport ideality factor $\nsig \neq 2$ for most of the range. Clearly, both ideality factors originate from the trapping and subsequent release of charge carriers within the active layer, and their analytical models depend on the density of localised states.\cite{hofacker2017dispersive,baranovskii2018mott,saladina2023power} To address this important detail, we have incorporated ideality factors into the analytical model of $\alpha$,
\begin{equation}\begin{split}\label{eq:alpha_long}
    \alpha
    &= \frac{eL}{\kT} \cdot \frac{j_{00}}{\sigma_{00}} \cdot \exp\bL -\frac{\eg - e\Voc}{\kT} \bl \frac{1}{\nid} - \frac{1}{\nsig}\br\bR \\
    &\propto \Phi^{1 - \nid/\nsig} . \\
\end{split}\end{equation}

The prefactor $j_{00}/\sigma_{00}$ determines the magnitude of $\alpha$. As mentioned in the previous section, recombination is generally dominated by the faster charge carrier, while the effective conductivity -- being a harmonic mean -- by the slower charge carrier. Hence, achieving a ratio of charge carrier mobilities that is close to unity, along with a lower Langevin reduction factor, becomes essential for decreasing the prefactor. This finding highlights the importance of balanced charge carrier mobilities.

The slope of $\alpha$ in Figure~\ref{fig:04}(a) corresponds to $1-\nid/\nsig$, in accordance with Eq.~\eqref{eq:alpha_long}. This ratio depends on which type of charge carrier, electrons or holes, dominates the transport and recombination processes. The analytical expressions for $\nid$ and $\nsig$ directly depend on the DOS these charge carriers occupy. We previously demonstrated that the DOS for electrons and holes in PM6:Y6 can be described by a combination of Gaussian and power-law state distributions, where the latter is approximated by an exponential function at a given QFLS.\cite{saladina2023power} Recombination in PM6:Y6 is primarily driven by mobile charge carriers from the Gaussian DOS interacting with charge carriers trapped in the power-law DOS. Charge transport takes place within the same density of states, therefore the transport ideality factor $\nsig$ is related to the recombination ideality factor $\nid$. When the same mobile charge carrier governs both effective conductivity and recombination, applying the multiple trapping and release model results in $1-\nid/\nsig = 0.5$ (see Supporting Information, Section~S6), meaning that $\alpha$ scales with $\sqrt{\Phi}$. On the other hand, if transport is limited by mobile charge carriers from the power-law DOS, then $1 -\nid/\nsig = 1.5 - \nid$, and $\alpha$ scales with a different power of light intensity. This power is equal to 0.5 only if $\nid=1$; in other cases, it is lower than 0.5. 

\begin{figure*}[t]\centering
    \includegraphics[width=0.32\textwidth]{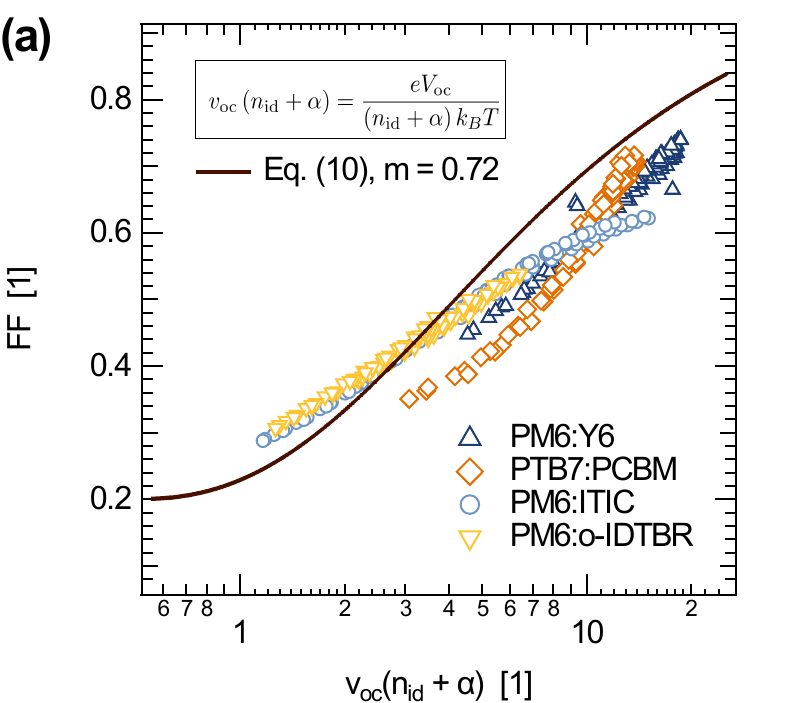}\quad
    \includegraphics[width=0.32\textwidth]{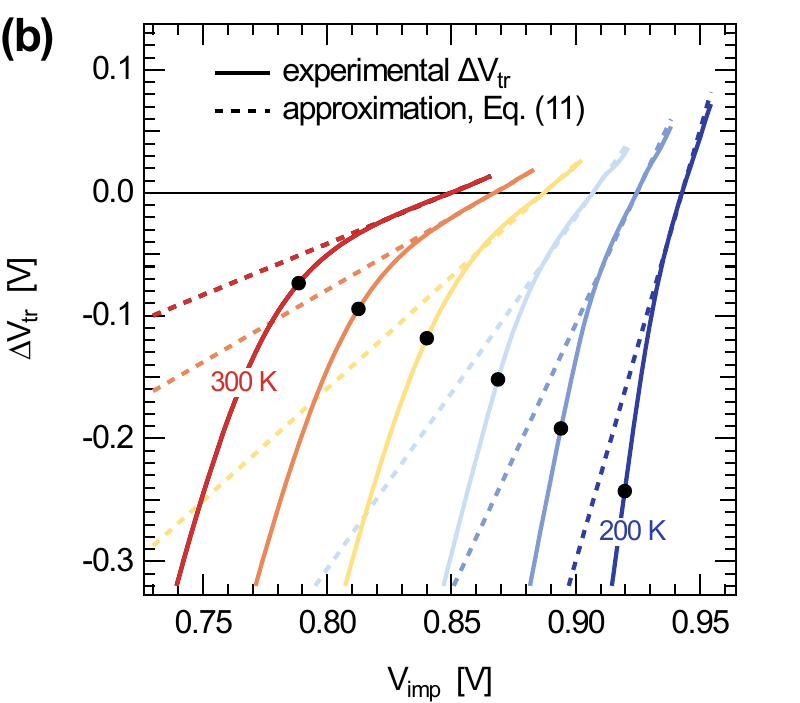}\quad
    \includegraphics[width=0.32\textwidth]{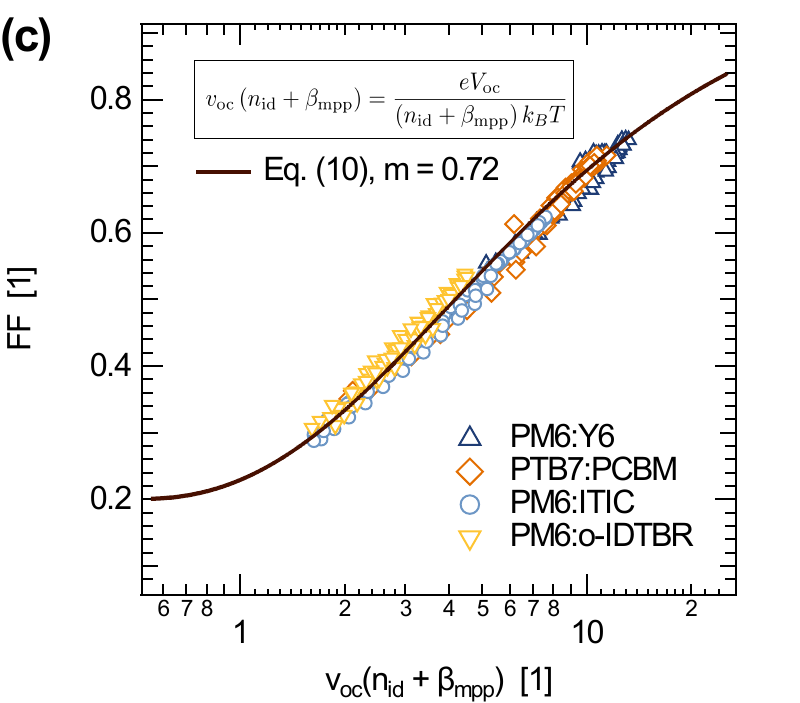}
    \caption{(a) The relationship between $\FF$ and $\voc$ varies depending on the material system, and cannot be described by the model, Eq.~\eqref{eq:FFgreen}. $\voc$ was determined for each material system using Eq.~\eqref{eq:FFvnorm} with $\napp = \nid+\alpha$. (b) Comparison between experimental and calculated $\Vtr$ for PM6:Y6. The approximation using $\alpha$ (Eq.~\eqref{eq:Vtr-approx}, dashed lines) underestimates the experimental $\Vtr$ (solid lines) at MPP (black dots). (c) The model given by Eq.~\eqref{eq:FFgreen} with $m=0.72$ applies universally to experimental $\FF(\voc)$. In contrast to (a), here experimental $\voc$ was evaluated using $\napp = \nid+\beta_\text{mpp}$, which accurately accounts for $\Vtr$ at MPP. }
\label{fig:05}\end{figure*}

The slope in Figure~\ref{fig:04}(a) is $<0.5$ at lower QFLS (higher $T$ and/or lower $\Phi$), aligning precisely with the ratio of the ideality factors depicted in Figure~\ref{fig:04}(c). This alignment indicates that in the PM6:Y6 solar cells under investigation, charge transport is limited by charge carriers from the power-law DOS. The correspondingly lower conductivity dominates the effective conductivity due to the harmonic mean. Conversely, in our earlier study,\cite{saladina2023power} we reported that \emph{recombination} in PM6:Y6 is primarily driven by mobile charge carriers from the Gaussian DOS that recombine with charge carriers trapped in the power-law DOS. Indeed, the data indicates that recombination and transport are governed by opposite charge carrier types. At a higher QFLS, the slope is $0.5$, but the dominance of charge carriers is unclear, as $\nid$ is close to unity. 

Overall, a higher ratio of $1-\nid/\nsig$ indicates similar effective disorder, leading to a decrease in $\alpha$ and an improvement in the fill factor. Generally, we can state that, alongside achieving low effective disorder in the active layer, resulting equal electron and hole conductivities are important to yield improved fill factors.

\subsection{The maximum power point: predicting the fill factor by an adapted figure of merit $\beta_\text{mpp}$.}

Let us now turn our attention to the maximum power point, where the fill factor is defined. To obtain a unified perspective, we consider how effectively the figure of merit $\alpha$ accounts for its behaviour. We will show that the approximations in $\alpha$ to predict the voltage loss due to transport resistance deviate away from $\Voc$. We will propose an improved figure of merit which we call $\beta_\text{mpp}$.

The analytical expression for the $\FF$ requires the definition of the normalised voltage.\cite{green1981solar} Generally, we state a normalised voltage at a certain point $i$, for example MPP, as:
\begin{equation}\begin{split}\label{eq:FFvnorm}
    v_i(\napp) &= \frac{eV_i}{\napp \kT } . \\
\end{split}\end{equation}
The normalised voltages at MPP and open circuit can be related as $\vmpp \approx \voc - \ln( \voc + 1 )$.\cite{taretto2013accurate} By using this approximation, the fill factor can be expressed using the normalised open-circuit voltage $\voc$ (for details, see Section~S8 in the Supporting Information)\cite{green1981solar,green1982accuracy,taretto2013accurate,wurfel2016physics}
\begin{equation}\begin{split}\label{eq:FFgreen}
    \FF &= \frac{\voc -\ln( \voc + m )}{\voc +1} . \\
\end{split}\end{equation}
This equation was used for inorganic solar cells to estimate an upper limit of the $\FF$, assuming infinite shunt and zero series resistance.\cite{green1981solar,green1982accuracy} To better align with experimental results, the value of $m=1$ under the logarithm was empirically replaced by $m= 0.72$.\cite{green1981solar}

We will evaluate the validity of Eq.~\eqref{eq:FFgreen} for organic solar cells. Initially, we will consider the effect of transport resistance at open circuit using the figure of merit $\alpha$. Later, we will present a more general view, as it becomes evident that predicting a parameter representing the MPP -- the $\FF$ -- cannot be accurately done using a figure of merit defined at open circuit -- namely $\alpha$.

We have assessed $\alpha$ experimentally for a wide range of temperatures and illumination intensities for four OSCs. Its relation to the $\FF$, shown in Figure~S5, indicates the direction for $\FF$ values at higher illumination intensities and increasing effective disorder, for example, by lowering the temperature (c.f.\ Figures~\ref{fig:03}(b) and \ref{fig:04}(a)) or through degradation within the device. To apply the expression describing the $\FF$, Eq.~\eqref{eq:FFgreen}, we calculate the normalised open-circuit voltage using $\napp=\nid+\alpha$ in Eq.~\eqref{eq:FFvnorm}. This approach was originally suggested for $\nid=1$;\cite{neher2016new} however, we use the measured values of $\nid$ instead. The results are shown in Figure~\ref{fig:05}(a). In qualitative terms, the model relating the experimental $\FF$ and $\alpha$ is highly effective -- but only for each material system individually. Consequently, the fitting equation for $\FF(\alpha)$ is unique to each system, while Eq.~\eqref{eq:FFgreen} does not apply to any of them.

The next aspect to unravel is why $\alpha$ cannot accurately predict the $\FF$ of all solar cells using a single equation. As already mentioned, $\nid+\alpha$ determines the slope of the $j(\Vext)$ curve at open circuit. $\Vtr$ can be approximated using a Taylor expansion at this point, as described in Section~S9. We find -- extending the work by Neher et al.\cite{neher2016new} -- that around $\Voc$:
\begin{equation}\begin{split}\label{eq:Vtr-approx}
    \Vtr &\approx \frac{\alpha}{\nid} (\Vimp-\Voc) . \\
\end{split}\end{equation}
This relationship determines the capability of the figure of merit $\alpha$ to predict the fill factor. It is shown in Figure~\ref{fig:05}(b) for a PM6:Y6 solar cell, alongside the measured $\Vtr$. In close proximity to $\Voc$, the approximation given by Eq.~\eqref{eq:Vtr-approx} demonstrates excellent agreement with the data. However, away from $\Voc$, the discrepancy between the experimental data and the model becomes more evident. Particularly, at the MPP (marked by black dots), the model significantly underestimates $\Vtr$.

The parameter $\alpha$ remains constant at a given temperature and light intensity; in other words, it does not depend on the implied and external voltage and is only valid at $\Voc$. Away from this point, transport resistance loss is more precisely described by the parameter $\beta$, as given by Eq.~\eqref{eq:beta}. Hence, an accurate prediction of the $\FF$ requires an apparent ideality factor $\nid+\beta$ evaluated at the MPP, rather than $\nid+\alpha$, which is evaluated at open circuit. As an effective way to determine $\beta_\text{mpp}$, in Section~S9 of the Supporting Information, we propose a fast-converging iterative scheme to determine this figure of merit from $\alpha$, $\Voc$, and the ideality factors $\nid$ and $\nsig$.

In Figure~\ref{fig:05}(c), we show that Eq.~\eqref{eq:FFgreen} -- with the normalised open-circuit voltage containing the apparent ideality factor $\napp = \nid+\beta$ -- is sufficient without modifications to describe the fill factor of all investigated solar cells in the temperature range of 200 to 300\,K. The parameter $\beta$ is defined by Eq.~\eqref{eq:beta} and is evaluated at the MPP. The validity of Eq.~\eqref{eq:FFgreen} is significant for several reasons. It demonstrates that the fill factor of OSCs is essentially determined by the open-circuit voltage of a solar cell, along with its recombination and transport resistance losses, represented by $\nid$ and $\beta_\text{mpp}$, respectively. Moreover, Eq.~\eqref{eq:FFgreen} allows these losses to be separated. The pseudo-fill factor of a solar cell containing only recombination losses can be estimated by setting $\beta_\text{mpp}=0$. In this case, the normalised open-circuit voltage is determined solely by recombination -- namely $\nid$. Next, we will present our perspective on how the framework of transport resistance allows to evaluate $\FF$ losses and explore potential strategies for their mitigation.

\subsection{Strategies to reduce $\FF$ losses}

To discuss ways to minimise fill factor losses due to transport resistance, we employ two types of metrics: the fill factor yield, $\eta_{\FF}$, and the collection efficiency at the maximum power point, $\eta_\text{col,mpp}$. These are valuable for considering losses at MPP from different perspectives.

\begin{figure}[b]\centering
    \includegraphics[width=0.38\textwidth]{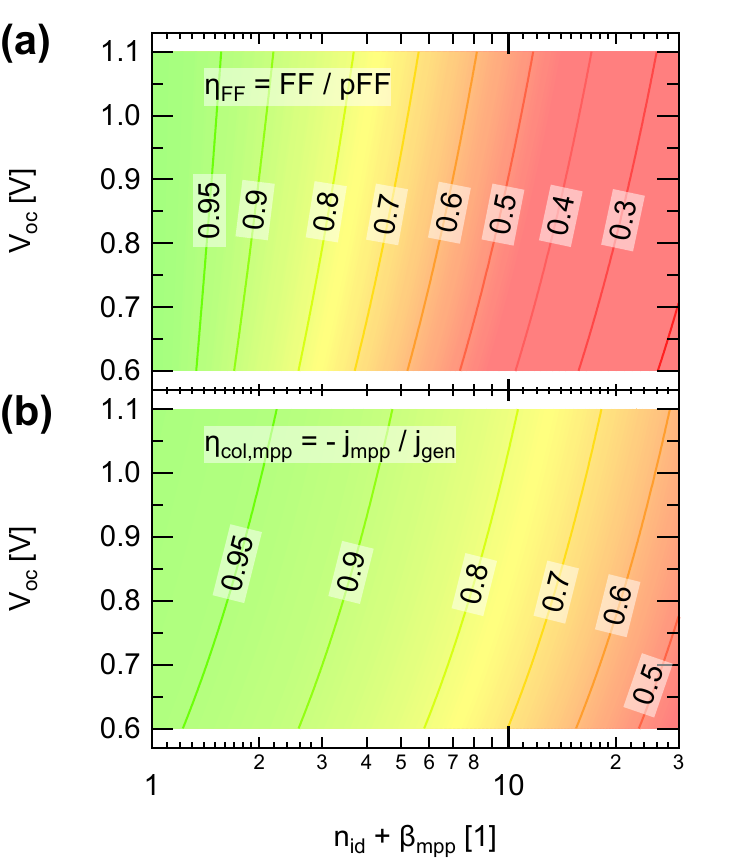}
    \caption{Calculated values of (a) the fill factor yield, $\eta_{\FF}$, using Eq.~\eqref{eq:eta-FF}, and (b) collection efficiency at the maximum power point, $\eta_\text{col,mpp}$, according to Eq.~\eqref{eq:eta-col}. Here, $\nid=1$ and $T=300$~K.}
\label{fig:06}\end{figure}

The fill factor yield is a measure of the $\FF$ loss due to transport resistance.
It relates the actual $\FF$ to its higher limit for a solar cell without transport resistance, the $\pFF$. To demonstrate the impact of transport resistance on this metric, we approximate Eq.~\eqref{eq:FFgreen} using a simple function $\FF \approx \voc/(\voc+4.37)$, similar to the approach by Green,\cite{green1982accuracy} which is valid for $\FF$ values between 0.4 and 0.9. Using this approximation, the $\FF$ yield is given by
\begin{equation}\begin{split}\label{eq:eta-FF}
    \eta_{\FF} = \frac{\FF}{\pFF} = \frac{e\Voc + 4.37\nid\kT}{e\Voc + 4.37\bl\nid+\beta_\text{mpp}\br\kT} .
\end{split}\end{equation}
When transport resistance is absent, $\beta_\text{mpp} = 0$, and $\eta_{\FF} = 1$. However, as the voltage loss due to transport resistance increases, $\eta_{\FF}$ tends to 0. In real devices, $\beta_\text{mpp}$ is always greater than 0 but can be minimised. The prediction of $\FF$ loss based on Eq.~\eqref{eq:eta-FF} is depicted in Figure~\ref{fig:06}(a). For a typical state-of-the-art organic solar cell with $\Voc$ of $0.85$~V, the predicted $\pFF \approx 0.87$, estimated from Eq.~\eqref{eq:FFgreen} assuming $\nid=1$. To attain a high fill factor of $0.8$, $\eta_{\FF}$ must exceed $0.9$. Therefore, maintaining $\beta_\text{mpp}<1$ is crucial for achieving such high fill factors. We note that, if both $\Voc$ and the short-circuit current density $\jsc$ remain unchanged, the power conversion efficiency without transport resistance loss would be higher by a factor of $1/\eta_{\FF}$ relative to the actual power conversion efficiency. 

Fill factor losses are also related to the collection efficiency, $\etacol$. It is quantified as the ratio of the collected current density $j$ to the current density generated within the active layer of the solar cell $\jgen$. Voltage loss due to the transport resistance is linked to the collection efficiency by
\begin{equation}\begin{split}\label{eq:Vtr-eta}
    \Vtr
    &= -\frac{\alpha\kT}{e} \cdot\etacol \cdot \etarec^{-\nid/\nsig} , \\
\end{split}\end{equation}
where $\etarec = 1-\etacol$ is the recombination efficiency, the complementary metric of $\etacol$. Eq.~\eqref{eq:Vtr-eta} demonstrates that the $\FF$ loss depends on the competition between charge carrier collection and recombination, a concept previously discussed by Bartesaghi et al.\ using the figure of merit $\Theta$.\cite{bartesaghi2015competition}

Since the voltage at MPP can be linked to $\Voc$, it allows the adoption of several useful simplifications when evaluating solar cell parameters. The collection efficiency at MPP can be simply expressed as
\begin{equation}\begin{split}\label{eq:eta-col}
    \eta_\text{col,mpp} &= \frac{\voc}{\voc+1} = \frac{e\Voc}{e\Voc + \bl\nid+\beta_\text{mpp}\br\kT} .
\end{split}\end{equation}
The results are shown in Figure~\ref{fig:06}(b), highlighting that minimising transport resistance also reduces recombination losses. Setting $\beta_\text{mpp}=0$ allows to estimate the collection efficiency in a transport resistance-free solar cell. Assuming that charge generation is field-independent, the internal quantum efficiency, $IQE\propto\etacol$, at MPP is mainly determined by $\voc$. Therefore, the collection efficiency at MPP, just as the voltage loss $\Vtr$, is directly influenced by energetic disorder: the ideality factors $\nid$ and $\nsig$ are functions of the DOS, as are the figures of merit for transport resistance.

To mitigate $\FF$ losses and improve collection efficiency, it is essential to minimise the apparent ideality factor $\nid + \beta_\text{mpp}$, as evident from Figure~\ref{fig:06}. Several strategies can be employed for this purpose. One approach involves reducing energetic disorder to attain $\nid=1$ and decreasing the trap density to minimise the rate of trap-assisted recombination. The difference in the effective energetic disorder of electrons and holes is influenced by the shape of the DOS and impacts the ratio between $\nid$ and $\nsig$. Values of the ratio exceeding its minimum value of $0.5$ -- corresponding to equal disorder (Section~S6) -- increase transport resistance losses: they raise $\beta_\text{mpp}$ compared to $\alpha$ (Eq.~(S4)), as well as $\alpha$ itself (Eq.~\eqref{eq:alpha_long}). Additionally, reducing the prefactor $j_{00}/\sigma_{00}$ plays a crucial role in minimising $\alpha$, as follows from Eq.~\eqref{eq:alpha_long}. When recombination and transport are governed by opposite charge carrier types, unbalanced mobilities increase this prefactor, leading to a lower $\FF$. Finally, we emphasise that transport resistance losses scale linearly with the thickness of the device, which makes strategies to minimise such losses particularly important for the industrial-scale production of OSCs.

\section{Conclusion}

In conclusion, we investigated the factors influencing the fill factor of organic solar cells. We evaluated the transport resistance losses in various solution-processed organic solar cells, employing current--voltage and open-circuit voltage measurements. We presented a precise method for determining the effective conductivity at open-circuit conditions, enabling the accurate evaluation of the figure of merit $\alpha$, a measure of transport resistance at $\Voc$. The experimental observations revealed a strong correlation between the fill factor and $\alpha$, highlighting that fill factor losses due to low conductivity in organic solar cells are a common issue and deserve more attention from the research community. Even in solar cells with comparatively low transport resistance ($\alpha\approx 1$), the fill factor loss is over 10\% compared to scenarios without transport resistance.

Based on extensive experimental data, we generalised the analytical model for transport resistance to account for energetic disorder. We did this by considering the voltage dependence of recombination and transport, by including the corresponding ideality factors. We extended the diode equation accordingly, allowing for the evaluation of transport resistance losses at the maximum power point. The refined analytical model serves as a powerful tool for predicting the fill factor of a solar cell, based on its open-circuit voltage. Additionally, we introduced a metric for quantifying fill factor losses and collection efficiency at the maximum power point, along with strategies for mitigating these losses, thus enabling the development of more efficient photovoltaic devices.

\section*{Acknowledgements}
{\small We thank the Deutsche Forschungsgemeinschaft (DFG) for funding this work (Research Unit FOR~5387 POPULAR, project no.~461909888). }

\section*{Competing interests}
{\small The authors declare no competing interests.}

\section*{Data availability}
{\small The data supporting the findings of this study is available from the corresponding author upon reasonable request. }

\bibliographystyle{apsrev4-2}
\bibliography{references}

\end{document}

\title{Transport resistance strikes back: unveiling its impact on fill factor losses in organic solar cells}

\author{Maria Saladina}
\affiliation{Institut für Physik, Technische Universität Chemnitz, 09126 Chemnitz, Germany}

\author{Carsten Deibel}
\affiliation{Institut für Physik, Technische Universität Chemnitz, 09126 Chemnitz, Germany}

\begin{center}
    \large
    Supporting Information
    \vspace{0.2cm}
\end{center}

\maketitle

\section{Experimental methods}\label{sec:S1}

\subsection{Device fabrication} 

The materials PM6, PTB7, ITIC, o-IDTBR, and Y6 were acquired from 1-Material Inc., while PCBM was obtained from Solenne BV, and used as received. The solutions for the active layer blends were prepared in the following manner:
\begin{enumerate}
    \item PM6:Y6, 1:1.2~w/w, 10~mg\,ml$^{-1}$ in chloroform with 0.5 vol.-$\%$ of 1-chloronaphthalene, stirred overnight at room temperature; 
    \item PM6:ITIC, 1:1~w/w, 10~mg\,ml$^{-1}$ in chloroform with 0.5 vol.-$\%$ of 1-chloronaphthalene, stirred overnight at room temperature;  
    \item PM6:o-IDTBR, 1:1~w/w, 10~mg\,ml$^{-1}$ in chloroform with 0.5 vol.-$\%$ of 1-chloronaphthalene, stirred overnight at room temperature;  
    \item PTB7:PCBM, 1:1.5~w/w, 25~mg\,ml$^{-1}$ in chlorobenzene with 3 vol.-$\%$ of DIO, stirred overnight at 60~$^{\circ}$C. 
\end{enumerate}
Pre-patterned indium tin oxide (ITO)-coated glass substrates underwent cleaning in an ultrasonic bath with detergent, acetone, isopropanol, and deionised water. Subsequently, they were exposed to low-pressure oxygen plasma for 5~min. A 35~nm layer of poly(3,4-ethylenedioxythiophene) polystyrene sulfonate (PEDOT:PSS, Clevios AI~4083, Heraeus Deutschland GmbH \& Co.\ KG) was spin-coated and annealed at 140~$^{\circ}$C for 10~min. The active layers of nonfullerene acceptor solar cells was spin-coated in a nitrogen-filled glovebox from blend solution at 3000~r.p.m., while PTB7:PCBM blend was spin-coated at 600~r.p.m. PM6:ITIC and PM6:o-IDTBR were annealed at 100~$^{\circ}$C for 10~min. The nonfullerene acceptor solar cells were finalised by depositing a 5~nm layer of bathocuproine (Ossila BV) and 100~nm of thermally evaporated Ag. For PTB7:PCBM, a 2~nm layer of Ca and 150~nm of Al were thermally evaporated on top of the active layer through a shadow mask with a base pressure below 10$^{-6}$~mbar.
\vspace{0.3cm}

\subsection{Current-voltage measurements}
The samples were excited using a continuous wave laser Omicron~LDM~A350, operating at a wavelength of 515~nm. The laser's output power, alongside Thorlabs neutral density filters controlled by Standa motorised filter wheels, allowed for modulation of illumination intensity. Throughout the measurement, a silicon photodiode continuously monitored the illumination intensity. The current output was measured with a Keithley~2634b source measure unit. Throughout the experiment, the sample was maintained within a Linkam~Scientific~LTS420 cryostat. This cryostat ensured low temperatures via a constant flow of liquid nitrogen using a Linkam~Scientific~LNP96-S liquid nitrogen pump and Linkam~Scientific~T96-S temperature controller.

\clearpage
\section{Constructing series resistance-free suns-$\Voc$ curves}\label{sec:S2}

\begin{figure}[h]
    \centering
    \includegraphics[width=0.4\textwidth]{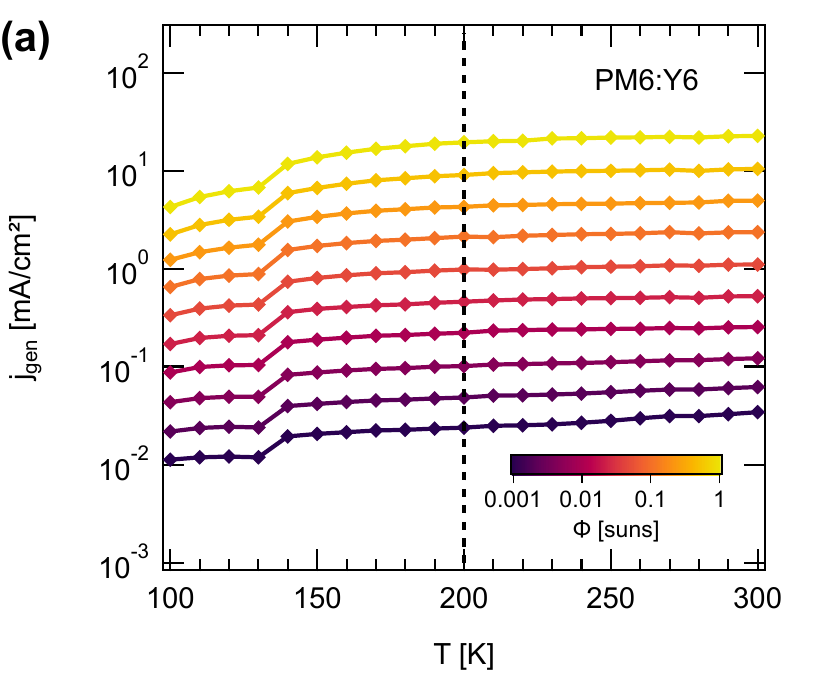}\quad
    \includegraphics[width=0.4\textwidth]{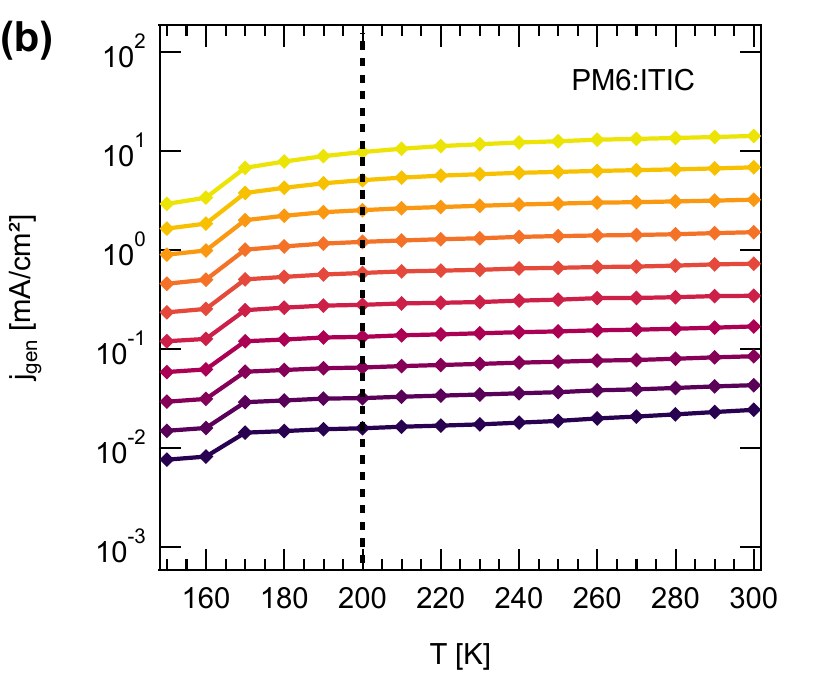}
    \\[\baselineskip]
    \includegraphics[width=0.4\textwidth]{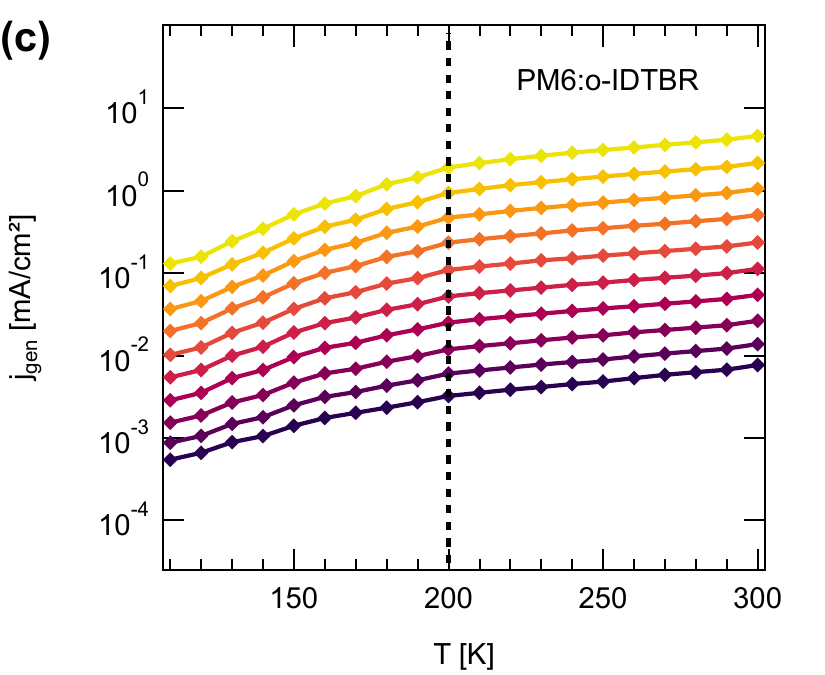}\quad
    \includegraphics[width=0.4\textwidth]{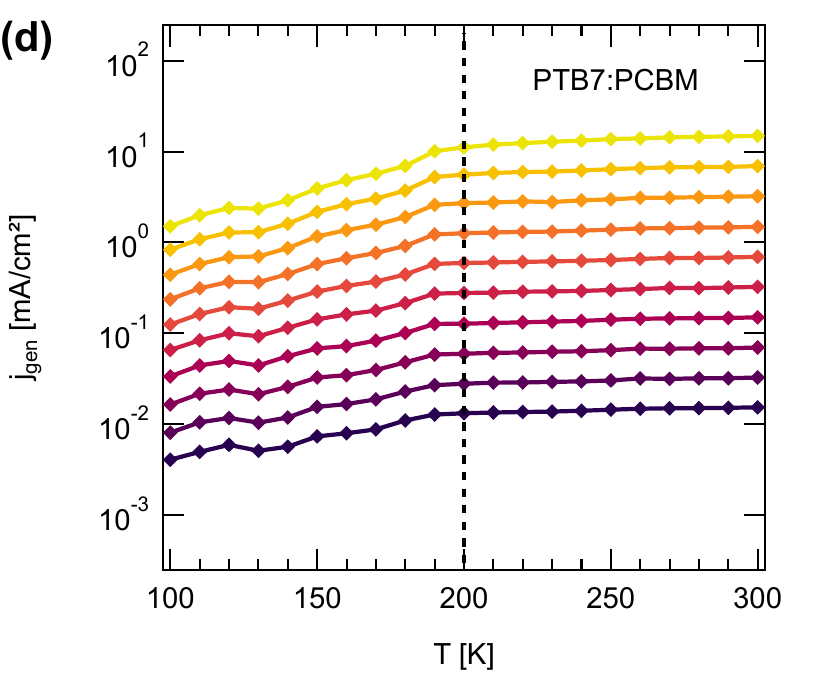}
    \caption{The generation current density, $\jgen$, which is estimated from the current density at -0.5~V, as a function of temperature for (a) PM6:Y6, (b) PM6:ITIC, (c) PM6:o-IDTBR, and (d) PTB7:PCBM. Above 200~K, indicated by the dashed line, $\jgen$ remains relatively constant. The data below 200~K was excluded from the analysis. }
    \label{fig:S01}
\end{figure}

The generation current density $\jgen$, shown in Figure~\ref{fig:S01}, is estimated from the current density at a negative bias of -0.5~V. 
At $\Voc$ it is equal to recombination current density, thus $\jrec(\Vimp)$ for a given temperature is constructed from $\jgen$ and $\Voc$ pairs at different light intensities. Finally, $j(\Vimp)$ at a given light intensity $\Phi$ is obtained by subtracting $\jgen(\Phi)$. 

\section{Influence of external series resistance on the modified diode equation}\label{sec:S3} 

External series resistance $\rext$ was determined using the derivative of external voltage with respect to current density. At high forward bias the transport resistance $\rtr$ is much smaller than $\rext$. Then 
\begin{equation*}\begin{split}
    j\bl V \br &\approx j_0\cdot \exp\bl \frac{eV - ej\rext}{\nid\kT} \br - \jgen \\
    V\bl j \br &\approx \frac{\nid\kT}{e}\ln\bl\frac{j+\jgen}{j_0}\br + j\cdot\rext \\
    \frac{dV\bl j \br}{dj} &\approx \frac{\nid\kT}{e} \cdot \frac{1}{j+\jgen} + \rext . 
\end{split}\end{equation*}

\begin{figure}[h]
    \centering
    \includegraphics[width=0.4\textwidth]{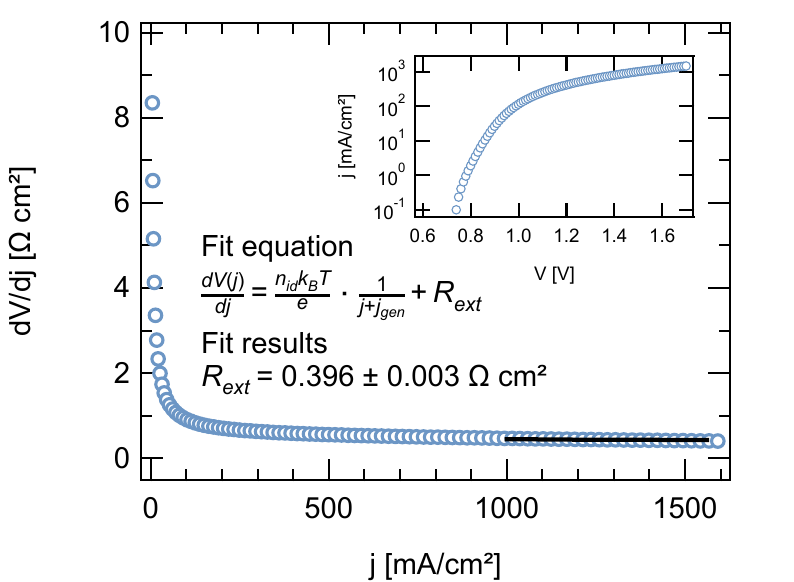}
    \caption{External series resistance $\rext$ for PM6:Y6, determined from the current--voltage characteristics at high forward bias, where the influence of $\rext$ is much higher compared to transport resistance. }
    \label{fig:S02}
\end{figure}

The fit is shown in Figure~\ref{fig:S02}. 
\vspace{0.3cm}

\begin{figure}[h]
    \centering
    \includegraphics[width=0.4\textwidth]{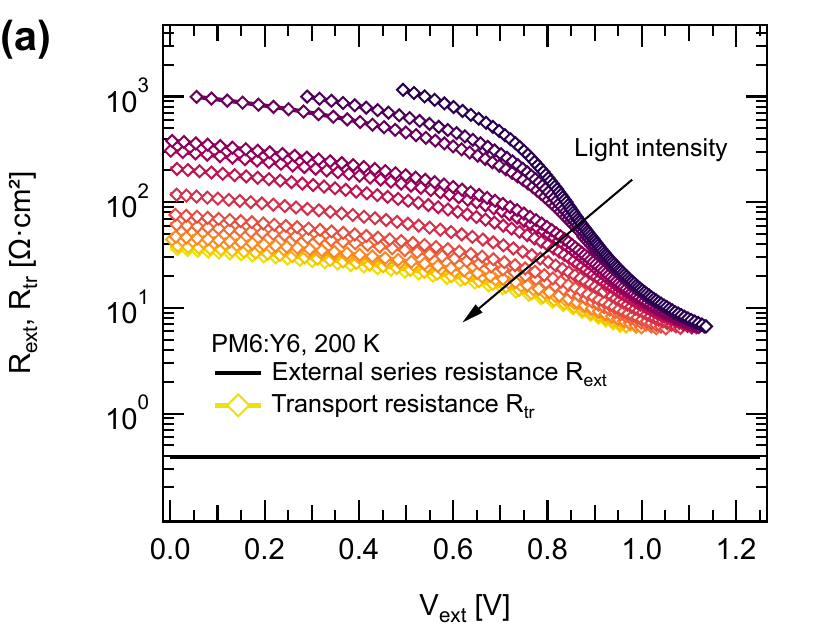}\quad
    \includegraphics[width=0.4\textwidth]{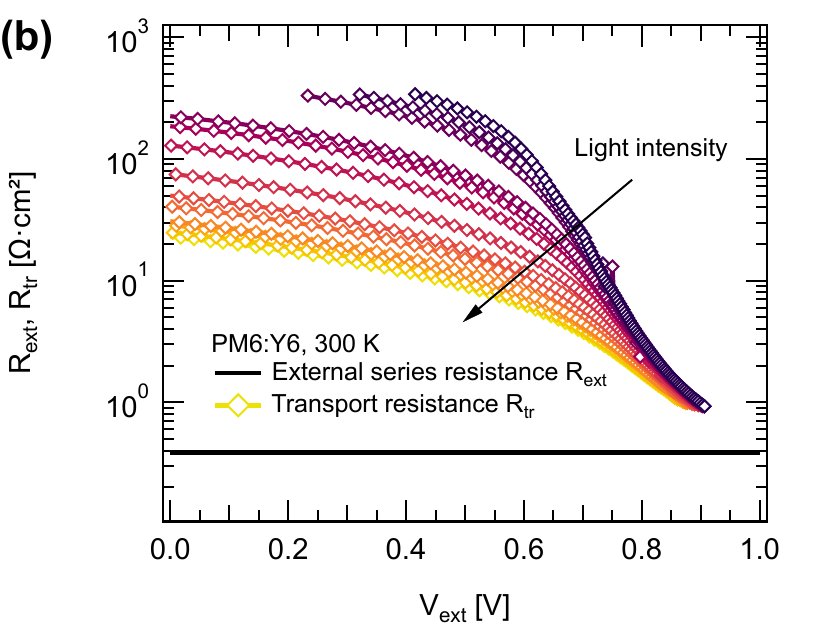}
    \\[\baselineskip]
    \includegraphics[width=0.4\textwidth]{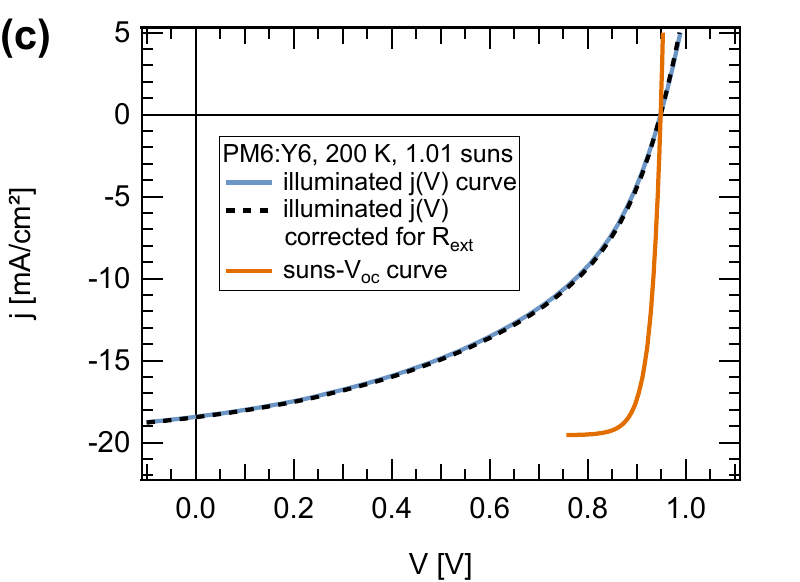}\quad
    \includegraphics[width=0.4\textwidth]{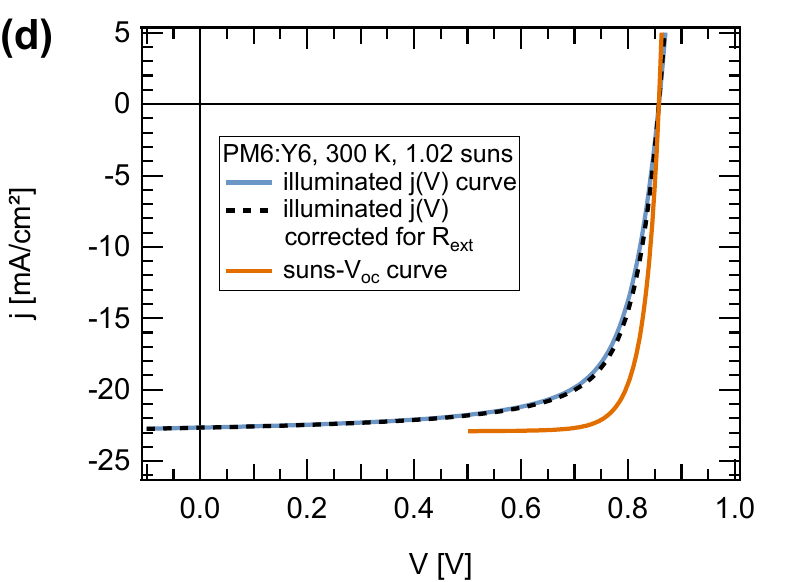}
    \caption{Transport resistance, $\rtr$ of PM6:Y6 compared to external series resistance $\rext$ at (a) 200~K, and (b) 300~K. The impact of $\rext$ for (c) 200~K, and (d) 300~K, is shown by the comparison of illuminated $j(V)$ curve, and its counterpart which takes into account the influence of $\rext$, i.e.\ the voltage is calculated according to $V = \Vext - j\cdot\rext$. The resistance free Suns-$\Voc$ curve is shown in order to highlight that the fill factor loss is mainly attributed to the transport resistance $\rtr$. }
    \label{fig:S03}
\end{figure}

While at high forward bias $\rext\gg\rtr$, the situation is opposite in the fourth quadrant of the $j(V)$ curve. Here $\rext$ has negligible impact on the $j(V)$ curve, as shown in Figure~\ref{fig:S03}.

\clearpage
\section{Influence of charge generation on the current--voltage characteristics}\label{sec:S4}

In this section, we extend Eq.~(4), which describes the current--voltage characteristics of an organic solar cell, to account for electric field-dependent charge generation. Charge generation is assumed to be governed by the competing processes of CT dissociation and geminate recombination \cite{braun1984electric}. By considering the coupled continuity equations for free charge carriers and CT states under steady-state conditions \cite{braun1984electric,gohler2018nongeminate}, the effective free charge carrier generation rate is given by
\begin{equation*}
    G_\text{eff} = \frac{k_d}{k_d + k_f} \cdot G_\text{CT} . 
\end{equation*}
Here, $G_\text{CT}$ is the generation rate of CT states, $k_f$ and $k_d$ are the geminate recombination and the CT dissociation rate constants, respectively. Generally, $k_d$ depends on the electric field, affecting the voltage dependence of $\jgen$. This voltage dependence can influence the $j(V)$ characteristics, the fill factor, and the figure of merit $\alpha$. 
\vspace{0.3cm}

We account for the relative impact of the electric field on $k_d$ as:
\begin{equation*}
    k_d(F) = k_d(0) \cdot \exp\bl \frac{eFr_0}{\kT} \br , 
\end{equation*}
where $k_d(0)$ is the dissociation rate constant at zero applied field, corresponding to the open-circuit conditions; $F$ denotes the electric field, and $r_0$ the separation distance between an electron and a hole in a CT state. This model implies that the efficiency of CT dissociation increases proportionally to the effect of the electric field on the electrostatic potential, and has been used previously for organic materials \cite{popovic1979electric,giebink2010ideal,saladina2021charge}. The electric field can be related to voltage as $F \approx (\Voc-\Vimp)/L$. Note that due to the effect of transport resistance on the external voltage, the latter is reduced and the electric field experienced by charge carriers is weaker. Hence, we use $\Vimp$ instead of the commonly used $\Vext$. For an intuitive explanation, see Figure~1 in reference~\intextcite{wurfel2015impact}. 
\vspace{0.3cm}

\begin{figure*}[t!]\centering
    \includegraphics[width=0.8\textwidth]{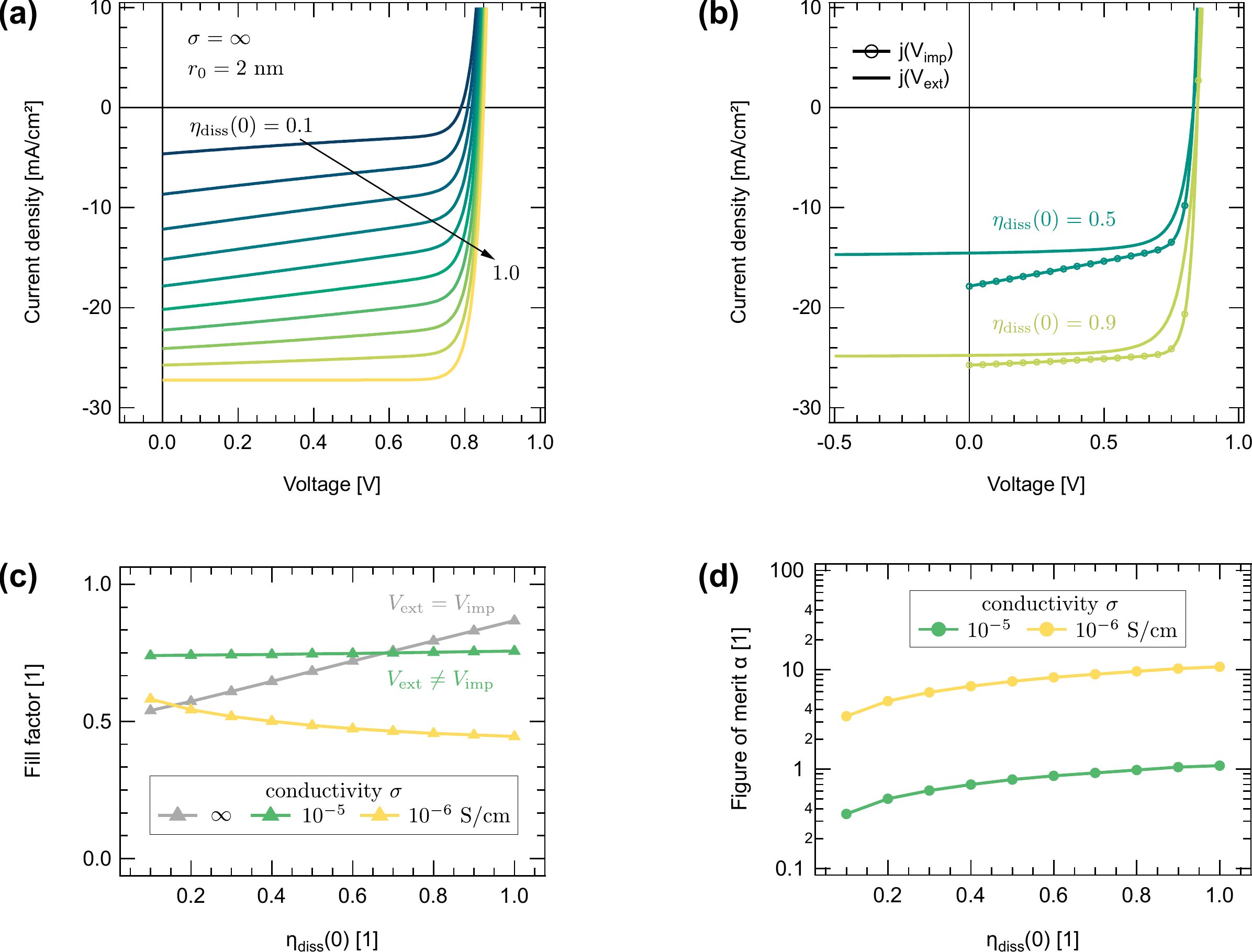}
    \caption{Simulated $j(V)$ curves with voltage-dependent charge generation. (a) For a hypothetical device with infinite conductivity. $r_0$ was set to 2~nm, with $\eta_\text{diss}(0)$ varying from 0.1 to 1. (b) $j(\Vimp)$ corresponds to a hypothetical device (infinite conductivity), while $j(\Vext)$ to a realistic device (finite conductivity). For the latter, the electric field in the fourth quadrant is much weaker due to the voltage loss caused by transport resistance. (c) $\FF$ as a function of $\eta_\text{diss}(0)$ for solar cells with varying $\sigma$. For realistic devices, $\FF$ is almost unaffected by charge generation for $\eta_\text{diss}(0) > 0.5$, while transport resistance has a clear impact. (d) The voltage-dependent charge generation minimally impacts the figure of merit $\alpha$ for $\eta_\text{diss}(0) > 0.5$, with transport resistance being the dominant factor impacting $\alpha$.}
\label{fig:S04}\end{figure*}

Using the above results, the generation current density is expressed as:
\begin{equation*}\begin{split}
    \jgen 
    &= eL G_\text{CT} \cdot \frac{k_d(F)}{k_d(F)+k_f} \\
    &= eL G_\text{CT} \bl 1 + \frac{k_f}{k_d(0)} \cdot\exp\bl \frac{e\Vimp - e\Voc}{(L/r_0)\kT} \br \br^{-1} . 
\end{split}\end{equation*}

To study the impact of $\jgen$ on the fill factor, we simulate $j(V)$-curves using Eq.~(4) from the main text, with $\jgen$ defined as outlined above. The variable parameter is the CT dissociation yield at zero field, defined as $\eta_\text{diss}(0) = k_d(0) / (k_d(0) + k_f)$, and representing the fraction of CT states dissociating under open-circuit conditions. First, we consider a solar cell unaffected by transport resistance losses, i.e., charge carriers are collected instantly after being generated with infinite conductivity. The parameters for calculation were chosen to match $\jsc$ and $\Voc$ of PM6:Y6 at 1~sun. The results are shown in Figure~\ref{fig:S04}. We fix $r_0$ at 2~nm in (a), and vary $\eta_\text{diss}(0)$ from 0.1 to 1.0. The slope of the $j(V)$-curve is affected both near short-circuit and open-circuit. These results indicate that for a \emph{hypothetical} solar cell, where \emph{conductivity is infinite}, electric-field dependent charge generation largely contributes to the fill factor losses. 
\vspace{0.3cm}

Let us consider a more realistic scenario, shown in Figure~\ref{fig:S04}(b), by examining a solar cell with finite conductivity ($\sigma_{\Voc} = 10^{-5}$~S~cm$^{-1}$, similar to the PM6:Y6 value at one sun illumination). Due to finite conductivity, the QFL gradient reduces $\Vext$, i.e., when 0~V is applied to the solar cell, the internal voltage $\Vimp$ remains close to $\Voc$. Consequently, the electric field inside the device is much weaker compared to the first example where $\Vext$ and $\Vimp$ were equal. This reduction in applied voltage manifests as a stretching of the $j(\Vimp)$-curve along the $x$-axis. As a result, the illuminated $j(V)$-curve, $j(\Vext)$ is minimally affected by field-dependent charge generation in the fourth quadrant with the field dependence becoming significant only at much higher reverse bias. Note that $\eta_\text{diss}(0) = 0.5$ corresponds to a very inefficient solar cell. While such devices are neither the focus of this paper nor of major research interest, it is important to highlight that even in these cases, the fill factor is negligibly impacted by charge generation.
\vspace{0.3cm}

To summarise the combined effect of charge generation and transport resistance on the fill factor, Figure~\ref{fig:S04}(c) compares devices with varying $\sigma$. In solar cells with infinite conductivity (no transport resistance, $\Vext = \Vimp$), the $\FF$ is strongly affected by geminate recombination. However, in \emph{real solar cells} the $\FF$ is not changing above $\eta_\text{diss}(0) = 0.5$ and is unaffected by $r_0$, as transport resistance losses dominate. 
\vspace{0.3cm}

Finally, let us estimate how charge generation impacts the figure of merit $\alpha$. Figure~\ref{fig:S04}(d) shows $\alpha$ as a function of the dissociation yield. Above $\eta_\text{diss}(0) = 0.5$, charge generation has virtually no influence on the figure of merit $\alpha$. We remind the reader that a solar cell with $\eta_\text{diss}(0) = 0.5$ is a relatively inefficient solar cell. Even for such devices, where geminate recombination is one of the dominant losses, the slope around $\Voc$ is dominated by the losses due to transport resistance.

\clearpage
\section{Apparent ideality factor at $\Voc$}\label{sec:S5}

As described in the main text, the apparent ideality factor is defined as $\napp = \nid + \beta$. At $\Voc$, parameter $\beta$ given by Eq.~(3) is indeterminate because both $\Vtr = jL/\sigma$ and $\Vimp - \Voc$ are equal to 0. To evaluate $\beta$ in the limit of $j\to 0$, we use L'Hopital's rule
\begin{equation*}\begin{split}
    \lim_{j\to 0}\beta 
    &= \nid \cdot \frac{d\Vtr}{dj}\bL\frac{d\bl\Vimp-\Voc\br}{dj}\bR^{-1} 
\end{split}\end{equation*}

The derivative of $\Vtr$ is taken using Eq.~(2) in the main text. Generally 
\begin{equation}\begin{split}\label{eq:S01}
    \frac{d\Vtr}{d j} 
    &= L\bl\sigma^{-1} + j\cdot \frac{d\bl\sigma^{-1}\br}{dj}\br. \\
\end{split}\end{equation}

The derivative of $\Vimp - \Voc$ can be found using Eq.~(1) in the main text, resulting in
\begin{equation*}\begin{split}
    \frac{d\bl\Vimp-\Voc\br}{dj} 
    &= \frac{\nid\kT}{e}\cdot\frac{1}{j+\jgen}. \\
\end{split}\end{equation*}
At open circuit conditions $j=0$, and $d\Vtr/dj$ in Eq.~\eqref{eq:S01} reduces to just one term, yielding $L/\sigma_{\Voc}$. The above equation is also simplified, leading to 
\begin{equation*}\begin{split}
    \lim_{j\to 0}\beta 
    &= \nid \cdot \frac{L}{\sigma_{\Voc}} \cdot \bL\frac{\nid\kT}{e}\cdot\frac{1}{\jgen}\bR^{-1} 
    = \frac{eL}{\kT}\cdot\frac{\jgen}{\sigma_{\Voc}} = \alpha . \\
\end{split}\end{equation*}
This result yields $\left.\napp\right|_{j=0} = \nid + \alpha$. 
\vspace{0.3cm}

\section{The ratio of the ideality factors describing the voltage dependence of recombination and transport}\label{sec:S6}

Earlier we have shown that recombination in PM6:Y6 is dominated by mobile charge carriers from the gaussian DOS recombining with the trapped ones from the power-law DOS.\cite{saladina2023power} The latter was approximated by an exponential DOS at a given energy, and therefore the apparent characteristic energy $\eu$ was also energy dependent. The recombination ideality factor was analytically described as 
\begin{equation*}\begin{split}
    \nid\bl E \br &= \frac{\eu(E) + \kT}{2\kT} . 
\end{split}\end{equation*}

The transport ideality factor, $\nsig$, which describes the voltage dependence of conductivity, can be derived using the multiple trapping and release model. Conductivity of electrons $n$ and holes $p$ is defines as
\begin{equation*}\begin{split}
    \sigma_n &= e\mu_{\text{eff},n}\cdot n = e\mu_{0,n} \cdot \theta_n \cdot n , \\
    \sigma_p &= e\mu_{\text{eff},p}\cdot p = e\mu_{0,p} \cdot \theta_p \cdot p , \\
\end{split}\end{equation*}
where $\mu_\text{eff}$ denotes the effective (charge carrier density dependent) mobility, $\theta$ is the trapping factor, i.e.\ the share of mobile charge carriers, and $\mu_0$ is their mobility. 
\vspace{0.3cm}

At open circuit conditions the densities of electrons and holes are equal, and can be expressed analytically at a given quasi-Fermi level splitting as\cite{hofacker2017dispersive,saladina2023power}
\begin{equation*}
    n = p = n_i \cdot \exp\bl \frac{e\Voc}{\eu(E) + \kT}\br , 
\end{equation*}
where $n_i$ is the intrinsic charge carrier concentration. 
\vspace{0.3cm}

The trapping factor generally depends on the density of states. To simplify derivation we assign the Gaussian DOS to electrons and the power-law DOS to holes. The results will have no loss of generality. With this assumption, the trapping factors are given by\cite{hofacker2017dispersive,saladina2023power}
\begin{equation*}\begin{split}
    \theta_n &= \exp\bl-\frac{s^2}{2(\kT)^2}\br , \\
    \theta_p &= N_0^{1-\lambda(E)} \cdot p^{\lambda(E)-1} . \\
\end{split}\end{equation*}
Here $s$ is the standard deviation of the Gaussian distribution, $N_0$ is the total density of states, and $\lambda(E)$ is defined as $\eu(E)/\kT$. 
\vspace{0.3cm}

Using the above two equations, the conductivity of electrons and holes becomes
\begin{equation*}\begin{split}
    \sigma_n &= e\mu_{0,n} \cdot \exp\bl-\frac{s^2}{2(\kT)^2}\br \cdot n 
    \propto \exp\bl \frac{e\Voc}{\kT}\cdot\frac{\kT}{\eu(E) + \kT}\br , \\
    \sigma_p &= e\mu_{0,p} \cdot N_0^{1-\lambda(E)} \cdot p^{\lambda(E)} 
    \propto \exp\bl \frac{e\Voc}{\kT}\cdot \frac{\eu(E)}{\eu(E) + \kT}\br . \\
\end{split}\end{equation*}

It follows that depending on the density of states, the voltage dependence of conductivity is expressed differently. As already mentioned, recombination in PM6:Y6 is governed by mobile charge carriers from the Gaussian DOS. If transport resistance is dominated by the same mobile charge carrier, the ratio of the ideality factors is 
\begin{equation*}\begin{split}
    \frac{\nid}{\nsig} &= \frac{1}{2} .  
\end{split}\end{equation*}
However, if transport resistance is governed by the mobile charge carrier of the opposite type, then the ratio has different expression
\begin{equation*}\begin{split}
    \frac{\nid}{\nsig} &= \frac{\eu(E)}{2\kT} = \nid - \frac{1}{2} .  
\end{split}\end{equation*}
The ratio of the ideality factors helps to distinguish which density of states limits the fill factor. Only in the special case of $\nid=1$ the latter equation yields the same result $1/2$, and the dominance can not be determined.

\section{The relationship between the figure of merit $\alpha$ and the fill factor}\label{sec:S7}

\begin{figure}[h]
    \centering
    \includegraphics[width=0.73\textwidth]{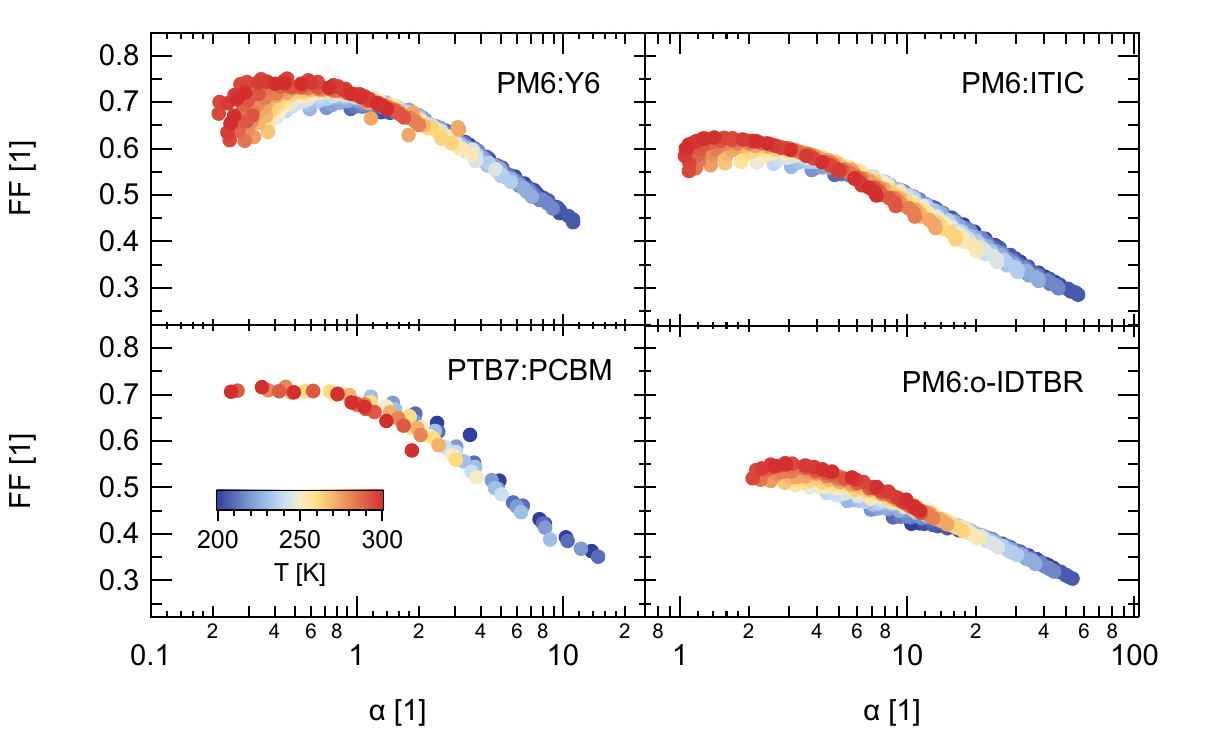}
    \caption{Fill factor as a function of $\alpha$ for PM6:Y6, PM6:ITIC, PTB7:PCBM, and PM6:o-IDTBR. The results validate the anticipated relationship between $\FF$ and $\ln\alpha$ as proposed by Neher et al.\cite{neher2016new} }
    \label{fig:S05}
\end{figure}

\clearpage
\section{Analytical approximation for the fill factor}\label{sec:S8}

The diode equation given by Eq.~(4) in the main text, can be rewritten using the normalised voltage $v_i$,\cite{green1981solar} as
\begin{equation}\begin{split}\label{eq:S02}
    j &= \jgen\bl \exp\bl \vext -\voc \br -1 \br, \quad \text{where} \quad v_i = \frac{eV_i}{\napp\kT}
\end{split}\end{equation}

At the maximum power point, the derivative of the $j\Vext$ product with respect to voltage is zero
\begin{equation*}\begin{split}
    0 &= j + \Vext\cdot\frac{dj}{d\Vext} \approx j + \vext\cdot\frac{dj}{d\vext} . 
\end{split}\end{equation*}
The latter expression can be verified using the chain rule, and it holds if $\napp$ changes little with voltage near the maximum power point. It leads to 
\begin{equation*}\begin{split}
    \exp\voc &= \exp\vmpp \cdot \bl \vmpp + 1 \br . 
\end{split}\end{equation*}

Using approximation of the Lambert W-function,\cite{taretto2013accurate} one finds that
\begin{equation*}\begin{split}
    \vmpp &\approx \voc - \ln\bl \voc + 1 \br . 
\end{split}\end{equation*}

Applying this result to Eq.~\eqref{eq:S02} for the current density yields
\begin{equation*}\begin{split}
    \jmpp &\approx \jgen\cdot\frac{\voc}{\voc+1} . 
\end{split}\end{equation*}
Note that in the latter expression $\jmpp$ is positive, therefore the minus sign is omitted. Finally, the fill factor is obtained using the last two equations, yielding Eq.~(10) in the main text.\cite{green1982accuracy} 
\vspace{0.3cm}

\begin{figure}[h]
    \centering
    \includegraphics[width=0.4\textwidth]{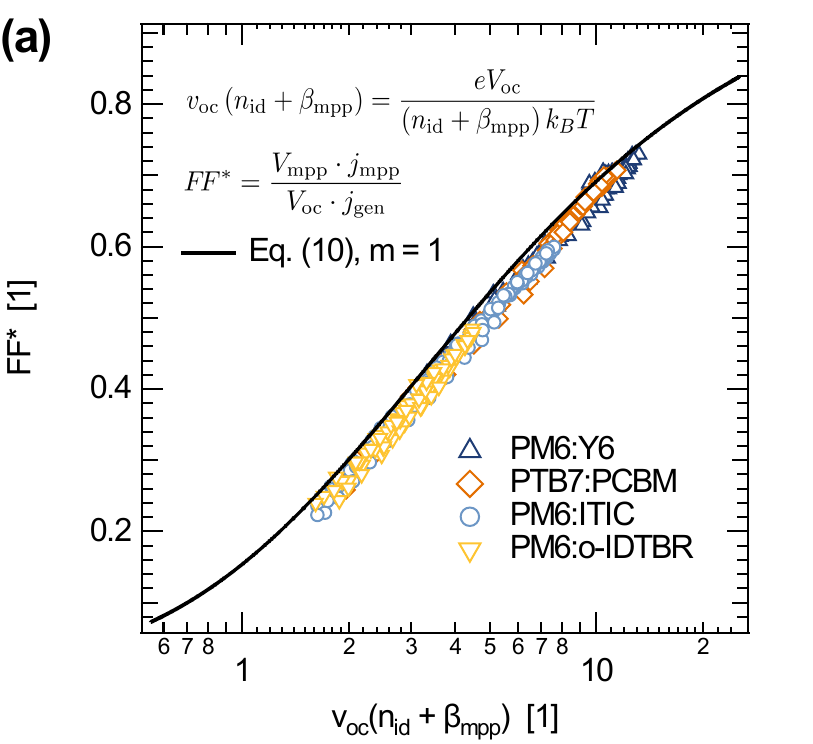}\quad
    \includegraphics[width=0.4\textwidth]{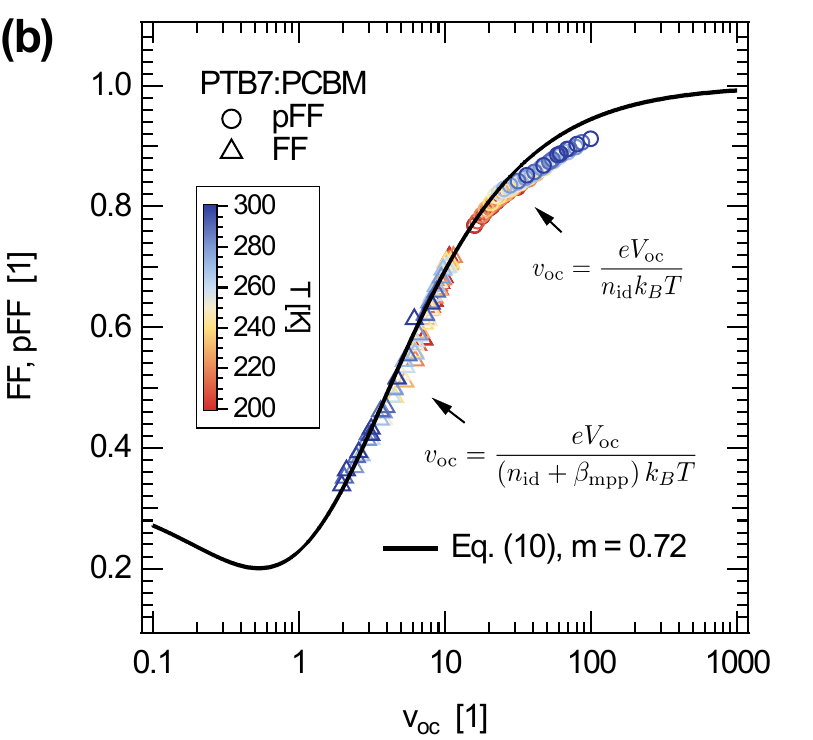}
    \caption{The fill factor approximation given by Eq.~(10) is applied to experimental data for different solar cells. (a) $\FF^*$ is calculated using $\jgen$ instead of $\jsc$. 
    (b) $\FF$ and $\pFF$ of a PTB7:PCBM solar cell. The pseudo-fill factor $\pFF$ represents the higher limit of $\FF$ obtained from the suns-$\Voc$ curve with zero transport resistance, therefore $\voc$ is determined by setting $\beta_\text{mpp}=0$. Eq.~(10) approximates the data in (a) with $m=1$, and in (b) with $m=0.72$. }
    \label{fig:S06}
\end{figure}

Interestingly, when Eq.~(10) is employed with $m=1$, it fits the fill factors as a function of normalised voltage when the former is evaluated using $\jgen$ instead of $\jsc$ (c.f.\ Figure~\ref{fig:S06}(a)). This adjustment aligns with the derivation of Eq.~(10), which requires the use of $\jgen$ in the diode equation. Green found empirically that setting $m = 0.72$ greatly improves approximation of the fill factor.\cite{green1981solar} Indeed, for the real fill factor calculated using $\jsc$, we find that Eq.~(10) with $m = 0.72$ works better, as shown in Figure~\ref{fig:S06}(b) and Figure~5(c) in the main text. Remarkably, Eq.~(10) with $m=0.72$ universally predicts the fill factor across values ranging from as low as 29~\% up to 75~\%. In Figure~\ref{fig:S06}(b) it is demonstrated to work reasonably well for fill factors at the higher limit, also describing the pseudo-fill factor in PTB7:PCBM.

\clearpage
\section{Voltage loss due to transport resistance}\label{sec:S9}

Current density is expressed via implied voltage $\Vimp$ using Eq.~(1) in the main text. Conductivity can be similarly expressed as a function of $\Vimp$
\begin{equation*}\begin{split}
    \sigma(\Vimp) &= \sigma_{\Voc} \cdot \exp\bl \frac{e\Vimp - e\Voc}{\nsig\kT} \br . 
\end{split}\end{equation*}

Derivation of $\Vtr$ is done in a similar way as in Neher et al.\cite{neher2016new} Substituting the above equation and Eq.~(1) into Eq.~(2) in the main text allows to obtain the general expression for the voltage loss due to transport resistance, $\Vtr$ 
\begin{equation}\begin{split}\label{eq:S03} 
    \Vtr(\Vimp) &= \frac{L\cdot j(\Vimp)}{\sigma(\Vimp)} \\
    &= \frac{\kT}{e} \alpha \left[ \exp\bl \bl 1 - \frac{\nid}{\nsig}\br \frac{e\Vimp - e\Voc}{\nid\kT}\br - \exp\bl -\frac{\nid}{\nsig}\cdot \frac{e\Vimp - e\Voc}{\nid \kT}\br \right] . \\
\end{split}\end{equation}

\subsection{$\Vtr$ at open circuit conditions}

Near the open circuit $\Vtr$ can be approximated using the Taylor expansion
\begin{equation*}\begin{split}
    \Vtr(\Vimp) &\approx \bl\Vimp-\Voc\br \cdot \left.\frac{d\Vtr}{d\Vimp}\right|_{\Vimp=\Voc} = \bl\Vimp-\Voc\br \cdot \frac{\alpha}{\nid} . 
\end{split}\end{equation*}

Consequently, $\alpha$ is a good measure of the transport resistance at these conditions. Away from $\Voc$ the exponential terms become sufficiently large, and the approximation deviates from the real value of $\Vtr$. The figure of merit $\alpha$ is not sufficient at the maximum power point to fully encompass $\Vtr$, and therefore $\beta$ has to be used.

\subsection{$\Vtr$ at the maximum power point}

Using Eqs.~(1) and (4) in the main text, we can replace $\Vimp$ with $\Vext$ in Eq.~\eqref{eq:S03}. We get
\begin{equation*}\begin{split}
    \Vtr 
    &= \frac{\kT}{e}\alpha \left[ \exp\bl \bl 1 - \frac{\nid}{\nsig} \br \frac{e\Vext - e\Voc}{\bl\nid+\beta\br\kT} \br - \exp\bl -\frac{\nid}{\nsig}\cdot\frac{e\Vext-e\Voc}{\bl\nid+\beta\br\kT} \br \right] \\
\end{split}\end{equation*}

Or, in terms of the normalised voltages
\begin{equation*}\begin{split}
    \Vtr 
    &= \frac{\kT}{e}\alpha \left[ \exp\bl \bl 1 - \frac{\nid}{\nsig} \br \cdot \bl \vext - \voc \br \br - \exp\bl -\frac{\nid}{\nsig}\cdot \bl \vext - \voc \br  \br \right]
\end{split}\end{equation*}

At the maximum power point, $\vext=\vmpp$, and $\vmpp-\voc \approx -\ln\bl \voc + 1\br$. With this approximation, the voltage loss due to transport resistance at MPP becomes 
\begin{equation*}\begin{split}
    \left.\Vtr\right|_\text{mpp} 
    &\approx \frac{\kT}{e}\alpha \left[ \exp\bl \bl 1 - \frac{\nid}{\nsig} \br \cdot \ln\bl \frac{1}{\voc+1} \br \br - \exp\bl -\frac{\nid}{\nsig}\cdot \ln\bl \frac{1}{\voc+1} \br \br \right] \\
    &= \frac{\kT}{e}\alpha \left[\bl \voc+1 \br^{\bl \frac{\nid}{\nsig}-1 \br} - \bl \voc+1 \br^{\frac{\nid}{\nsig}} \right] \\
    &= - \frac{\kT}{e}\alpha \cdot \voc \cdot \bl\voc+1\br^{\frac{\nid}{\nsig}-1} \\
\end{split}\end{equation*}

\begin{figure}[h]
    \centering
    \includegraphics[width=0.4\textwidth]{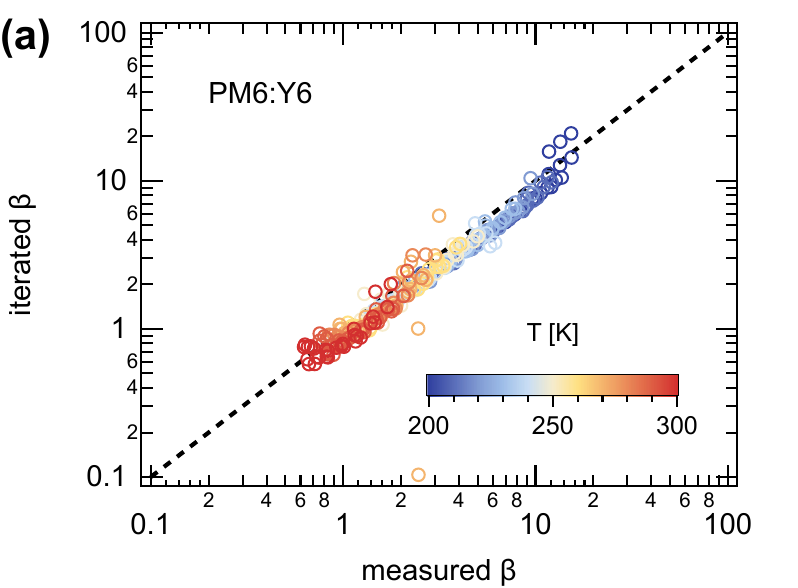}\quad
    \includegraphics[width=0.4\textwidth]{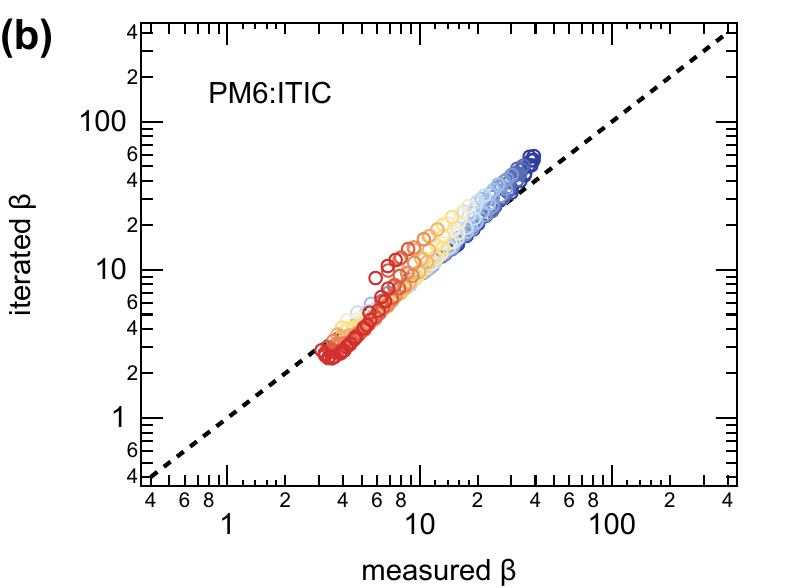}
    \\[\baselineskip]
    \includegraphics[width=0.4\textwidth]{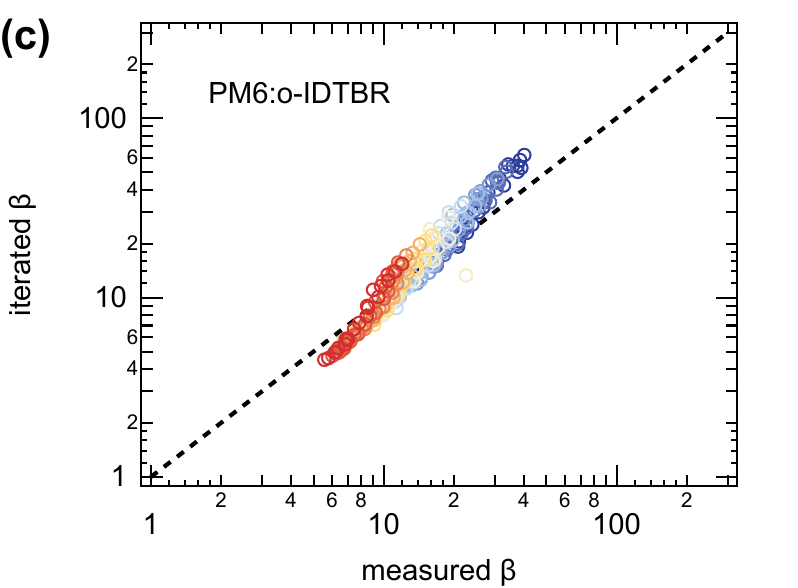}\quad
    \includegraphics[width=0.4\textwidth]{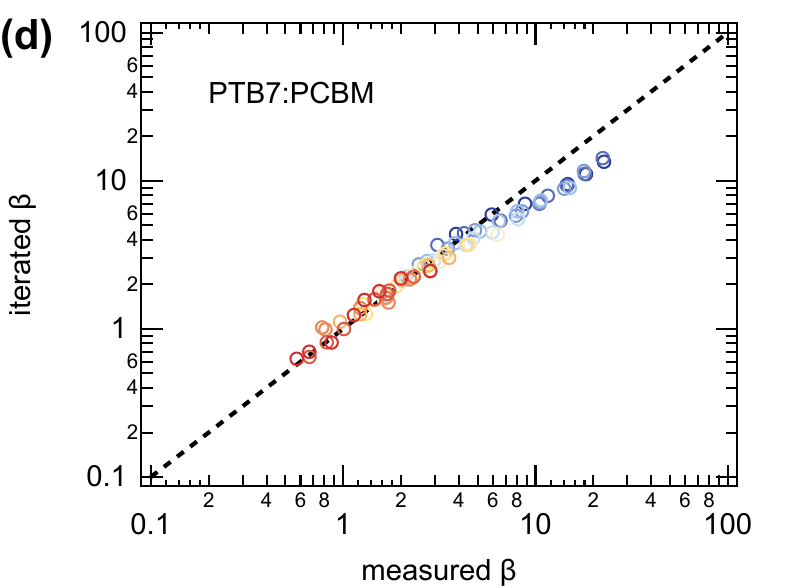}
    \caption{The parameter $\beta$ at the maximum power point determined through iteration using Eq.~\eqref{eq:S04}, and compared to measured $\beta$ for (a) PM6:Y6, (b) PM6:ITIC, (c) PM6:o-IDTBR, and (d) PTB7:PCBM. The dashed line indicates the equality between the iterated and actual values. }
    \label{fig:S07}
\end{figure}

Inserting this result into Eq.~(3) in the main text yields $\beta$ at MPP
\begin{equation}\begin{split}\label{eq:S04}
    \beta_\text{mpp} &= \alpha \cdot \frac{\voc \cdot \bl\voc+1\br^{\frac{\nid}{\nsig}-1}}{\ln\bl\voc+1\br} , \qquad \voc = \frac{e\Voc}{\bl\nid+\beta_\text{mpp}\br\kT} . \\
\end{split}\end{equation}
Consequently, at MPP $\beta$ can be predicted just from the values of $\alpha$, $\Voc$ and the ideality factors by iteration. This fast converging iterative scheme agrees reasonably well with the measured $\beta_\text{mpp}$ values, as shown in Figure~\ref{fig:S07}. Eq.~\eqref{eq:S04} helps to make predictions about the $\FF$ of a solar cell just from the physical parameters of the active material. 

\bibliographystyle{apsrev4-2}
\bibliography{references}